\documentclass{article}

\usepackage{titling}
\title{Transferable Knowledge for Low-cost Decision Making in Cloud Environments}

\newcommand*{\thename}{\texttt{{Tamakkon}}\@\xspace}
\newcommand{\fig}[1]{Figure~\ref{#1}}

\usepackage[sc]{mathpazo} 
\usepackage[T1]{fontenc} 
\usepackage{microtype} 
\usepackage[hmarginratio=1:1,top=25mm,columnsep=20pt]{geometry} 
\usepackage[hang, small,labelfont=bf,up,textfont=it,up]{caption} 
\usepackage{booktabs} 
\usepackage{float} 
\usepackage{paralist} 

\usepackage{abstract} 

\usepackage{titlesec} 
\renewcommand\thesubsection{\Roman{subsection}} 
\titleformat{\section}[block]{\large\scshape\centering}{\thesection.}{1em}{} 
\titleformat{\subsection}[block]{\large}{\thesubsection.}{1em}{} 

\usepackage{fancyhdr} 
\usepackage{amsfonts,amsmath,amssymb,amstext,latexsym}
\usepackage{color}
\usepackage[usenames,dvipsnames,table]{xcolor}
\usepackage{multirow}

\usepackage{graphicx}
\usepackage[caption=false]{subfig}

\usepackage{tabularx}
\usepackage[usenames,dvipsnames]{xcolor}
\newcolumntype{a}{>{\columncolor{CornflowerBlue!50}}c}
\usepackage{colortbl}
\usepackage{booktabs}
\newcommand{\midsepremove}{\aboverulesep=0mm \belowrulesep=0mm}
\midsepremove
\newcommand{\midsepdefault}{\aboverulesep=0mm \belowrulesep=0mm}
\midsepdefault

\usepackage[ruled]{algorithm2e}

\usepackage{framed}
\definecolor{mytab}{gray}{0.1}
\definecolor{myshade}{gray}{0.9}
\newenvironment{mybox}{%
  \vspace{-6pt}
  \MakeFramed{\advance\hsize-\width\FrameRestore}%
  \noindent%
  \itshape%
}
{\endMakeFramed}

\newcommand*{\sect}[1]{Section~\ref{#1}}
\usepackage{xspace}

\newcommand*{\eg}{\textit{e.g.}\@\xspace}
\newcommand*{\ie}{\textit{i.e.}\@\xspace}
\makeatletter
\newcommand*{\etc}{%
    \@ifnextchar{.}%
        {etc}%
        {etc.\@\xspace}%
}
\newcommand*{\etal}{%
    \@ifnextchar{.}%
        {et al}%
        {et al.\@\xspace}%
}
\makeatother

\begin{document}

\date{}
\title{\thetitle}

\author{
	Faiza Samreen, Gordon S Blair, Yehia Elkhatib\\[4mm]
    \small School of Computing and Communications, Lancaster University, United Kingdom\\
    \normalsize Email: \{\textit{i.lastname}\}@lancaster.ac.uk\\[4mm]
}

\maketitle

\thispagestyle{fancy} 

\begin{abstract}
Users of cloud computing are increasingly overwhelmed with the wide range of providers and services offered by each provider. As such, many users select cloud services based on description alone. An emerging alternative is to use a decision support system (DSS), which typically relies on gaining insights from observational data in order to assist a customer in making decisions regarding optimal deployment or redeployment of cloud applications. The primary activity of such systems is the generation of a prediction model (\eg using machine learning), which requires a significantly large amount of training data. However, considering the varying architectures of applications, cloud providers, and cloud offerings, this activity is not sustainable as it incurs additional time and cost to collect training data and subsequently train the models. We overcome this through developing a Transfer Learning (TL) approach where the knowledge (in the form of the prediction model and associated data set) gained from running an application on a particular cloud infrastructure is transferred in order to substantially reduce the overhead of building new models for the performance of new applications and/or cloud infrastructures. In this paper, we present our approach and evaluate it through extensive experimentation involving three real world applications over two major public cloud providers, namely Amazon and Google. Our evaluation shows that our novel two-mode TL scheme increases overall efficiency with a factor of 60\% reduction in the time and cost of generating a new prediction model. We test this under a number of cross-application and cross-cloud scenarios.
\end{abstract}

\section{Introduction}\label{sec:introduction}
\begin{figure}[hbt]
\centering
\includegraphics[width=0.5\columnwidth,clip,trim=0 0.25cm 0 0]{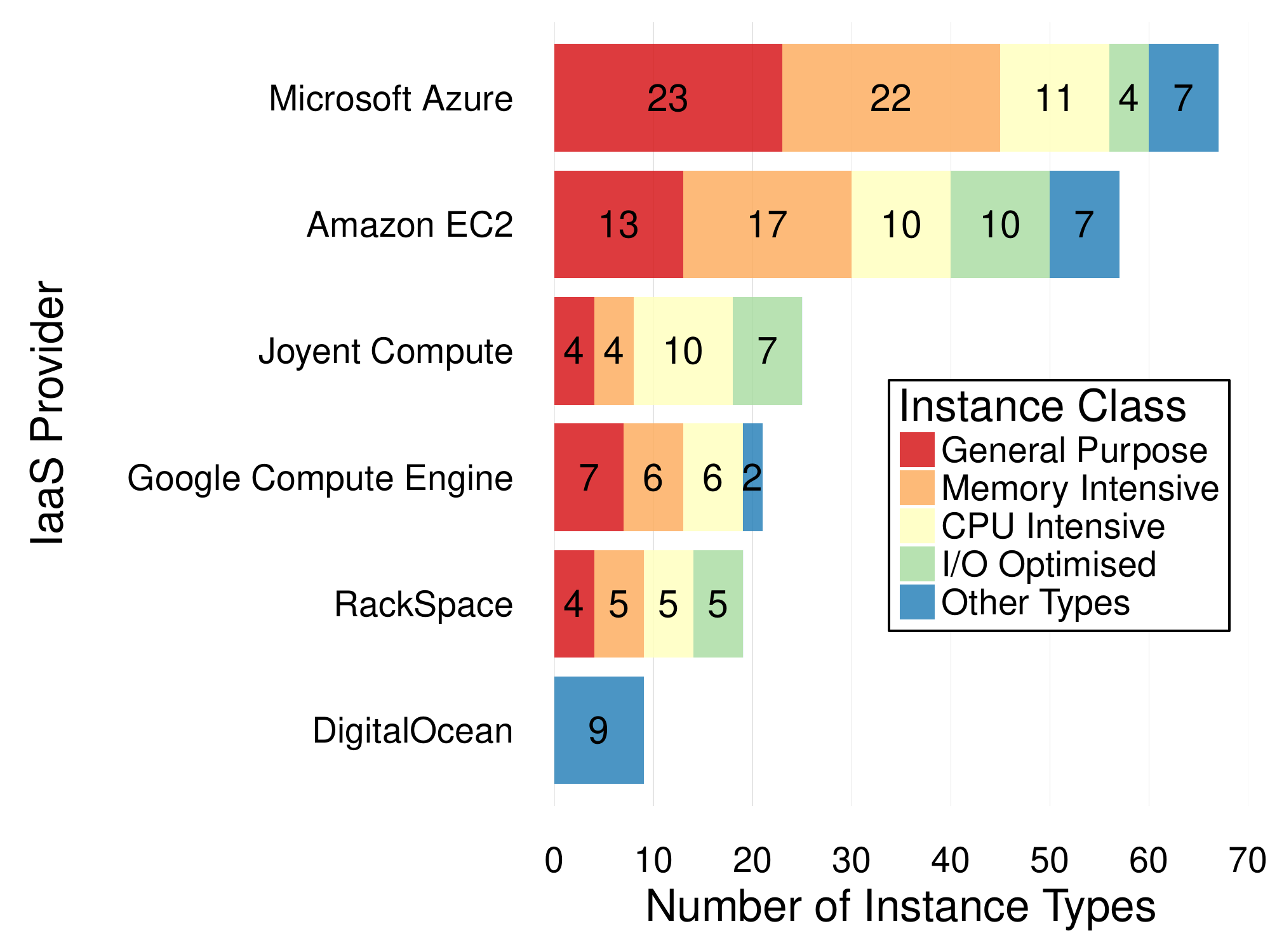}
\caption{On-demand instance types (Linux) offered by major IaaS vendors.}
\label{fig:instance-types}
\end{figure}

The cloud computing market has a proliferation of service offerings, pricing models, and technology standards~\cite{forbes2014choice,elkhatib2016Crosscloud}. 
This complicates decisions making regarding service selection~\cite{Zhonghong2012,Kilcioglu2017}. Although such challenges apply to all levels of cloud services, the Infrastructure as a Service (IaaS) level is particularly difficult given the fact that IaaS provides more choices and control for developers. In the IaaS domain, there is a wide range of virtual machines (VM) on offer -- see \fig{fig:instance-types} -- but no straightforward method to compare their performance and, more generally, cost/performance trade-offs, neither \textit{within} nor \textit{across} cloud providers. A wrong or suboptimal decision can result in financial loss as well as reduced application performance~\cite{Zhonghong2012,daleel,Ghrada2018cloudvnf}, a common concern of end users~\cite{etsi-csc-wp1,Lang2016,appdirect2017}.

\begin{figure*}[bh]
\centering
\begin{tabularx}{\textwidth}{@{}ccc@{}}

\includegraphics[width=0.32\textwidth, trim=0cm 0cm 0cm 0cm, clip]{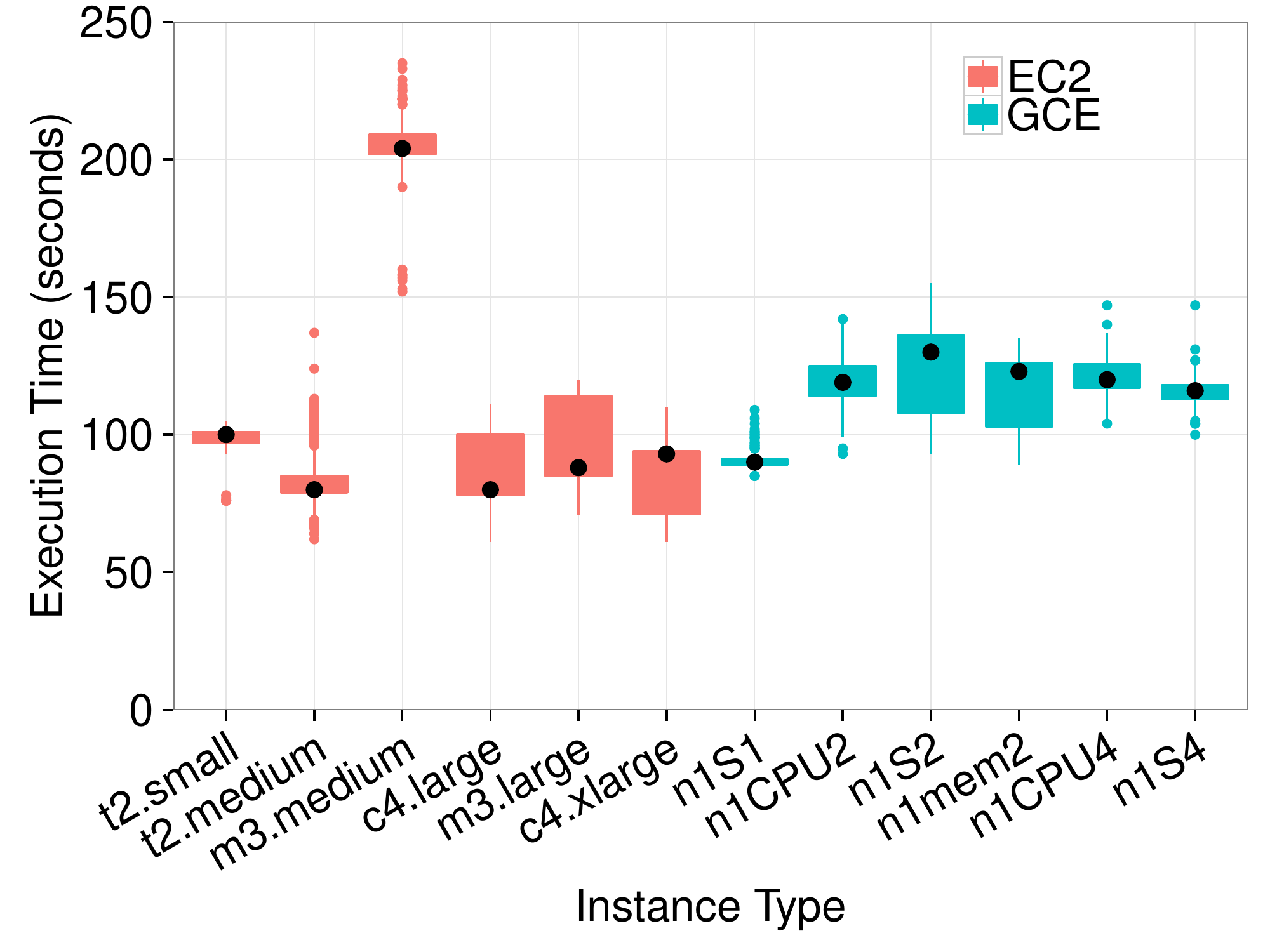}%
&
\includegraphics[width=0.32\textwidth, trim=0cm 0cm 0cm 0cm, clip]{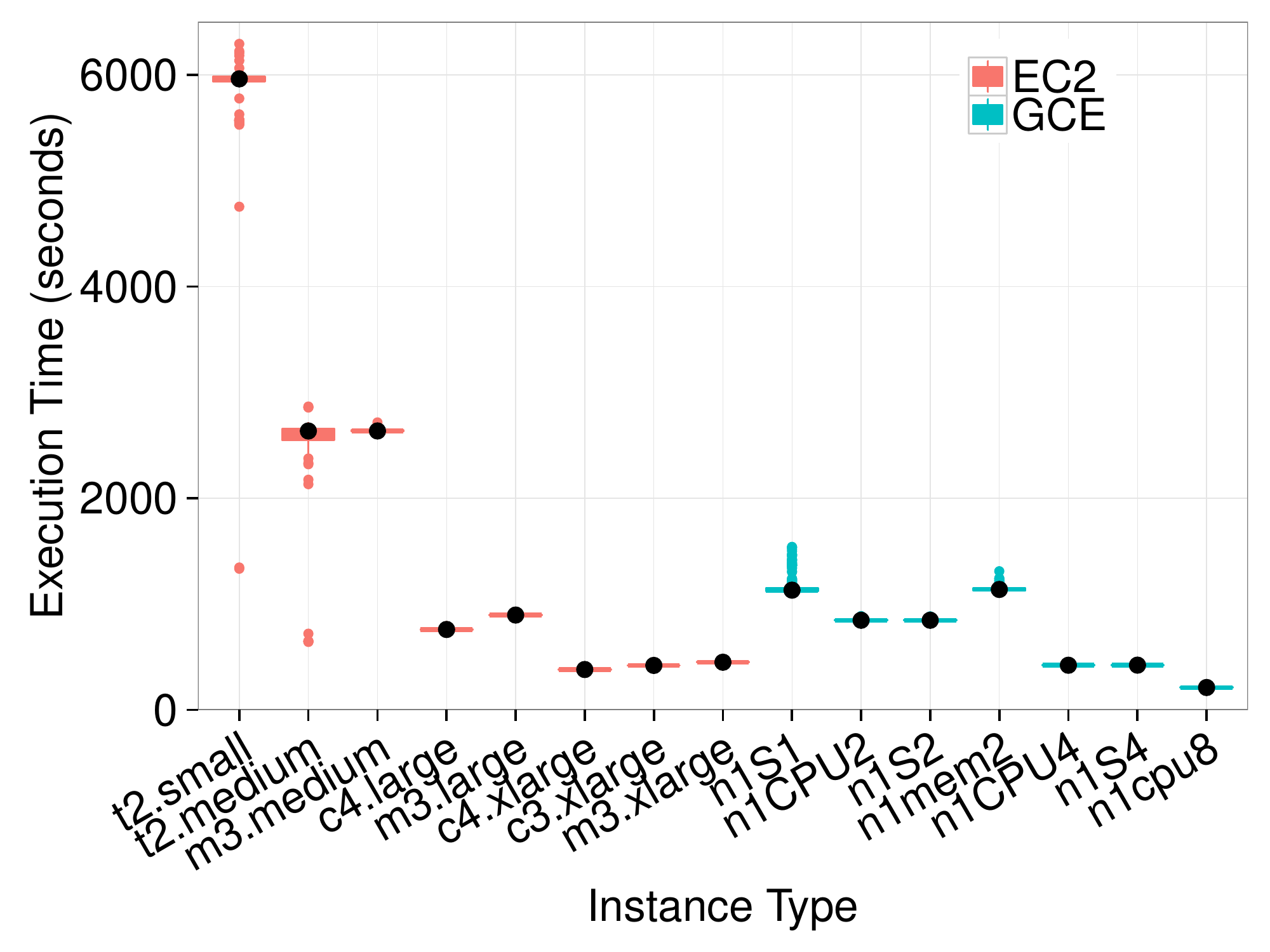}%
&
\includegraphics[width=0.32\textwidth, trim=0cm 0cm 0cm 0cm, clip]{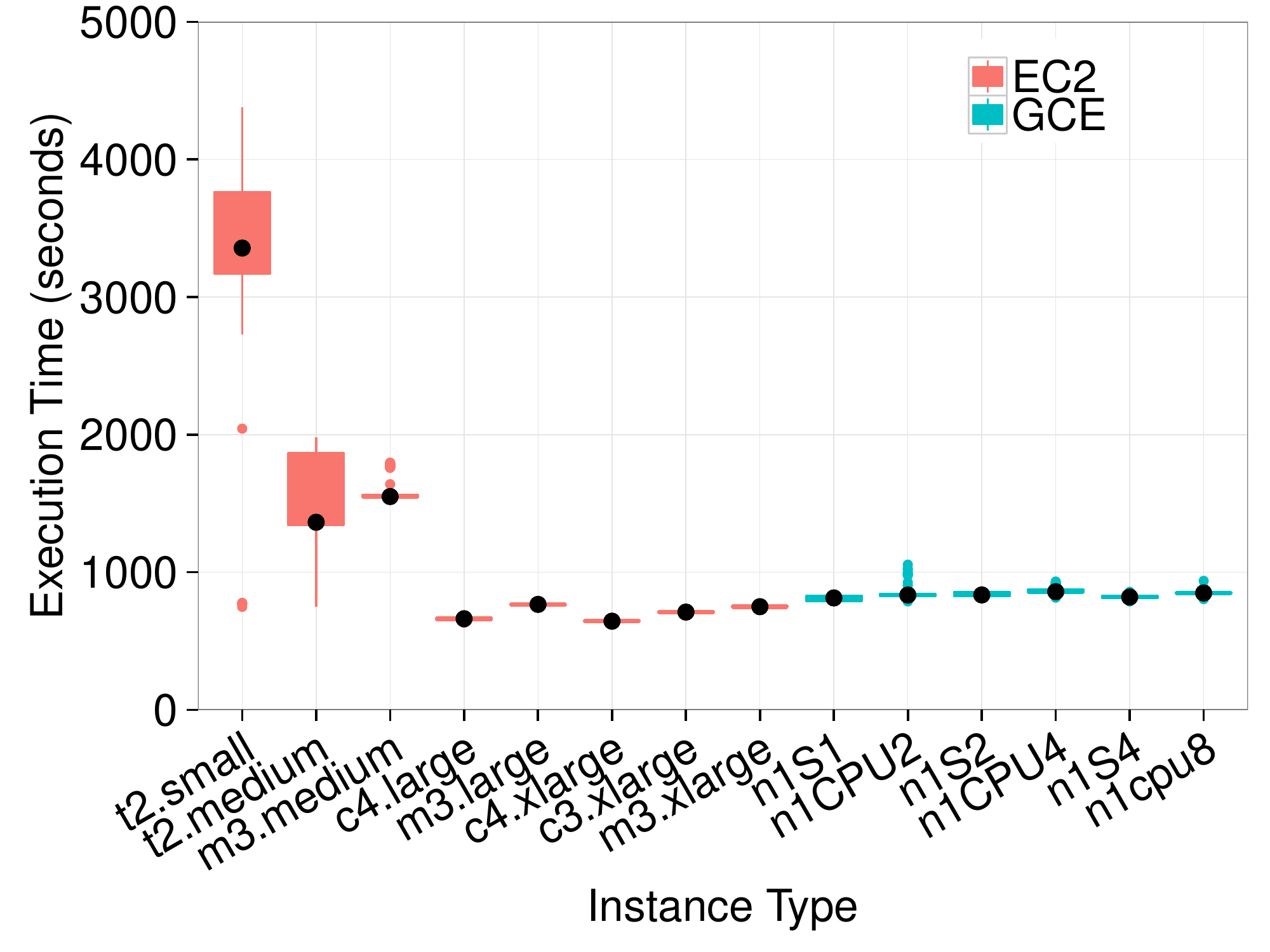}%
\\
(a) & (b) & (c)
\end{tabularx}
\caption{The execution times on various VM instances of AWS and Google of: (a) a memory-intensive application; (b) a CPU-intensive application; and (c) another CPU-intensive application. Instances from each provider are sorted from left to right in increasing instance hourly cost.}
\label{fig:execution-time-on-EC2-GCE}
\end{figure*}

The box plots in \fig{fig:execution-time-on-EC2-GCE} represent the distributional spread of execution times of 3 different applications across a variety of instance types of Amazon EC2 and Google GCE. \fig{fig:execution-time-on-EC2-GCE}(c), of a CPU-intensive application, reflects the least variability in performance levels and seems more predictable for GCE instances. The same pattern is observed across EC2 instances except the T2 series showing the highest execution time across all other nodes. 

For this CPU-intensive application, \texttt{c4.xlarge} (\$0.232/h), \texttt{c3.xlarge} (\$0.239/h) and \texttt{m3.xlarge} (\$0.280/h) with variable prices are performing at a very similar level, as shown in \fig{fig:execution-time-on-EC2-GCE}(b). Moreover, GCE's \texttt{n1cpu4} and \texttt{n1s4} show similar performance to the equivalent EC2 instances, and \texttt{n1cpu8} (\$0.215/h) is performing better than the expensive and predictable performing EC2 \texttt{m3.xlarge} (\$0.280/h). On the other hand, \texttt{c4.large} (\$0.116/h)is performing better than \texttt{c3.xlarge} at a much lower cost.

\fig{fig:execution-time-on-EC2-GCE}(c) reflects the same trend for the T2 series as seen in \fig{fig:execution-time-on-EC2-GCE}(b). A consistent performance with a negligible different in median values is observed among all GCE instances. 
In other words, the selection of a more expensive option does not lead to improved performance in terms of execution time.

\fig{fig:execution-time-on-EC2-GCE}(a) illustrates performance variability in executing a memory-intensive application on different instance types of variable computational capacity. Contrary to previous results, the T2 series is shown as the most predictable with less variability in terms of performance. In comparison with GCE, \texttt{n1s1} shows less variance in performance  and offers the best execution time compared to all GCE nodes as well as a most expensive EC2 node (\texttt{c4.xlarge}). On top of that, \texttt{n1S1} (\$0.036/h) and T2 instances (\$0.026-\$0.052/h) are the cheapest of all. Finally, a striking observation is the poor performance of \texttt{m3.medium} for both executions.

The following observations can be drawn from the above discussion. The selection of an instance type based on descriptive information cannot guarantee the best fit in terms of cost and performance trade-offs.
The distributional spread of results highlights the intrinsic uncertainty related to instance performance, contributing to the complexity regarding the selection of suitable instance. There are some external factors that can influence this performance such as the scheduling policy, machine age, provider-specific incentives etc. Therefore, nodes with similar capacity/specifications are difficult to compare.

Classified categories of instance types do not reflect the real performance behaviour of any application due to different architectures and the unique workload patterns of that application. Similarly, superficial knowledge about the application cannot guarantee the expected performance over different instance types.

In such decision-making situations, machine learning (ML) has potential to underpin intelligent and data-grounded Decision Support Systems (DSS). Indeed, ML-aided DSS have been demonstrated to provide behavioural and performance insights about application and deployment setup necessary to make optimal decisions, \eg~\cite{daleel,Chiang2014Matrix,Gencer2015ResponseSurface,AlicherryPick2017}. A traditional ML approach follows the general steps of: collecting data, generating a learning model, fitting the model on training data, and assessing its accuracy on test data. This is a well established, albeit lengthy, process. 
However, a common assumption in this traditional learning setting is that the test and training data sets are drawn from the same distribution (say, performance of a certain cloud provider's VMs using a particular application). If the distribution changes, such assumption no longer holds and, consequently, there needs to be a lengthy process of rebuilding the model. In our context, this means commissioning a new set of benchmarks that costs time and money. 

This presents a significant challenge for ML-based decision support in multi-cloud environments, where different architectures and varying deployment setups across different cloud providers lead to significant changes in data distribution as well as feature space. Therefore, in a real world scenario, conducting such experiments is time-consuming and requires considerable cost and time for data collection and model generation. 

This paper focuses on enhancing the learning efficiency of ML-aided DSS to assist application-specific decisions across different cloud providers.
In this paper, we propose a novel Transfer Learning (TL) assisted scheme leading to substantial reduction in the training overhead by making use of existing knowledge. Quantitatively, we observed a reduction of approximately 60\% in learning time and cost by transferring existing knowledge about an application and/or a cloud platform in order to learn a new prediction model for another application or cloud provider. The proposed scheme has the potential to deal with the challenge of model generation when data distribution or feature spaces differ between source and target domains.

 \subsubsection*{Contributions}
Our main contributions are as follows:
\begin{itemize}
\item We design and implement \thename, a two-mode TL scheme that identifies the type of knowledge to be transferred across domains. We utilize a subset of well-established machine learning approaches, namely Multiple Polynomial Regression (MPR), Support Vector Regression (SVR) and Random Forests (RF). We believe this to be a novel application of TL for decision support in cloud computing.

\item We present a methodology to measure similarity among different data sets in order to identify the source application and its learned data that can be used as a source of knowledge for generating a prediction model for a target application, and, additionally, to avoid negative knowledge transfer.

\item Through extensive experimentation, we apply \thename on different applications \textit{and} cloud providers. We efficiently generate a learning model between varying application and IaaS domains, reducing the learning cost (of cloud resources and time) by 60\%.
\end{itemize}

\subsubsection*{Novelty against the state of the art}
\thename has many unique features making it different from other similar research \cite{Yadwadkar2017paris, AlicherryPick2017, Venkataram16Ernest}. First of all, \thename is based on transfer learning approach and makes use of the existing model as well data to reduce the cost and time involved for model generation and large-scale data collection. Secondly, \thename works across different applications as well as cloud providers. Furthermore, a range of models are offered and the generated models result in reduced prediction errors. Moreover, the framework is extensible and new models can be added as base learners, if needed.

\section{Background on Transfer Learning (TL)}
\label{sec:bck}
\begin{figure}[htb]
\centering
\includegraphics[width=0.8\columnwidth]{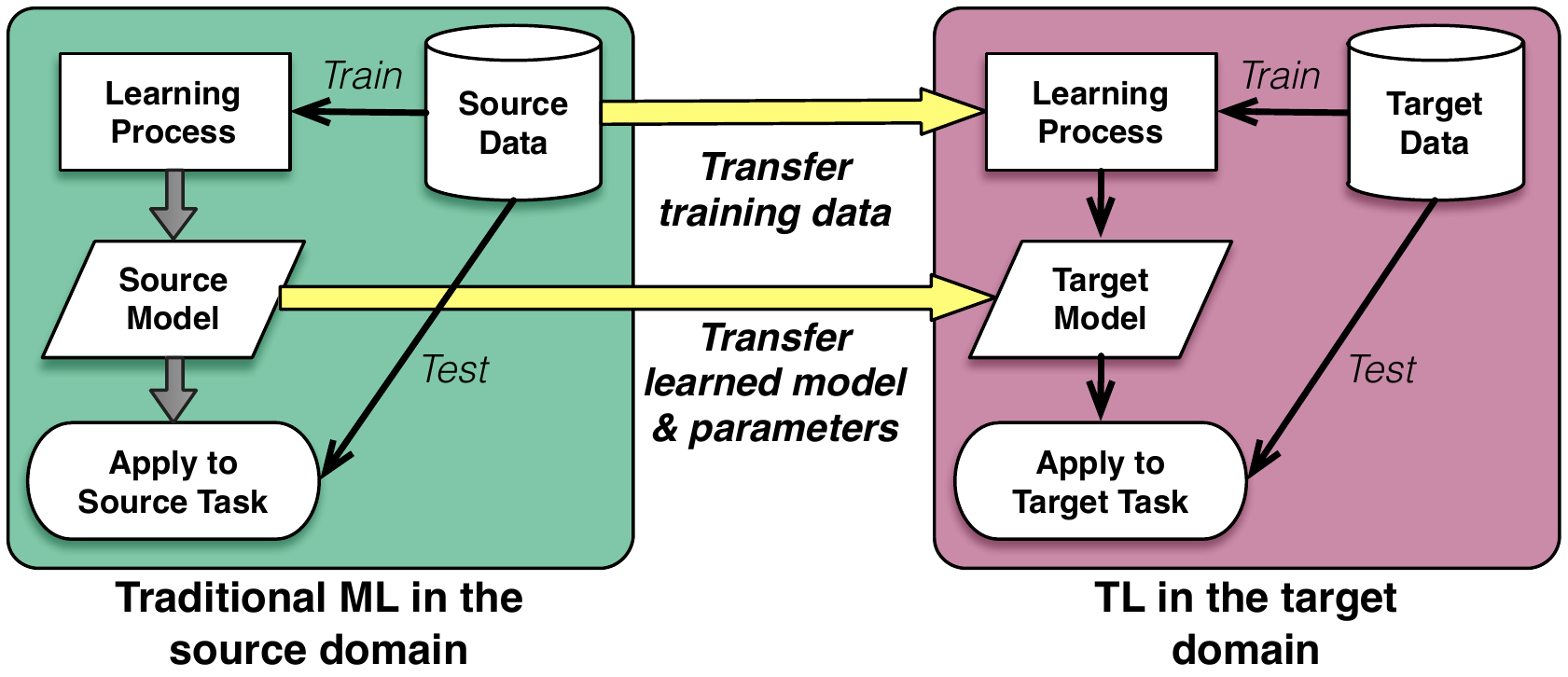}
\caption{An overview of the TL process.}
\label{fig:TL-arch}
\end{figure}

The general goal of TL is to use previously learned knowledge to solve a problem in an unseen environment or in a faster or better way. 
TL extracts knowledge from one or more source domains and source tasks, and applies that knowledge to achieve a task in a target domain. 
Given a source domain $D_S$ and source task $T_S$, a target domain $D_T$ and target task $T_T$, TL aims to improve the prediction function of $D_T$ using the knowledge in $D_S$ and $T_S$ considering that $D_S$ has some similarity with $D_T$ and $T_S = T_T$ or $T_S \neq T_T$. 
This process of achieving the target task by learning from the source domain and source task is illustrated in \fig{fig:TL-arch}. 

TL is categorized into two sub-types: 
\textit{inductive TL} aims to improve the learning of a target predictive function with the help of a source task, considering that the source and target domains are the same but the tasks differ; 
\textit{transductive TL} aims to improve the learning of a target predictive function with the help of knowledge from a source domain, where source and target domains may or may not be different but the tasks are the same. 

The literature offers different approaches of transferring knowledge between two domains~\cite{Pan2010SurveyTL}: The instance knowledge transfer approach is applied by re-weighting some portion of source data to be used in the target domain and iteratively measuring the model fitness for target task learning \cite{Liao2005InstanceAuxili,Dai2007BoostingTLInstance}. 
The feature representation knowledge transfer approach requires identification of good features that can reduce the differences between source and target domains in order to minimize model error and domain divergence \cite{Raina2007Selftaught,Argyriou2008Feature}. 
The parameter knowledge transfer approach is applied with an assumption that the source and target tasks share some parameters or prior distributions of the hyper-parameters of the models \cite{Lawrence2004Param,Bonila2007Param}. 
The relational knowledge transfer approach deals with TL problems in the relational domain. 
TL has been applied to many real-world applications, mostly in the fields of classification of images, verbal sentiment, software defects, as well as medical tomography~\cite{CheplyginaPPLSB17}. 
TL for regression tasks is a less researched area, however, some notable efforts have been made, 
\eg~\cite{Pardoe2010RegTL,Garcke2014RegTL,ChoiFSC17RegTL}. 

\section{\thename: A TL-assisted DSS}
\label{sec:design}

Our goal is to enhance the efficiency of intelligent DSSs in order to make data-driven and application-specific decisions in cross-cloud environments, \ie those spanning more than a single provider. 
We believe that efficiency can be achieved by reducing the training overhead through making use of existing knowledge in the form of experiential data sets, significant predictive features, and prediction models and their parameters. 
We devise a TL-based approach to assist in efficient model generation by transferring learned knowledge from a related domain. 
The proposed approach is designed specifically for deployment and migration decisions in at the IaaS level. 

Our approach is a semi-supervised transductive TL technique that allows the contribution of auxiliary target data for model generation. 
Semi-supervised learning is used due to its ability to learn with a little amount of labeled data, reducing the required amount of training data for the target domain, which is one of the key concerns of model generation efficiency. 

We now present the design and implementation of our solution (\S\ref{sec: System architecture}), and detail the algorithms underpinning the TL scheme (\S\ref{sec:similarity measure}--\ref{sec:2TL}).

\subsection{System architecture}
\label{sec: System architecture}

\begin{figure}[th]
\centering
\includegraphics[width=0.7\columnwidth,clip,trim=0.1cm 0 0.2cm 0]{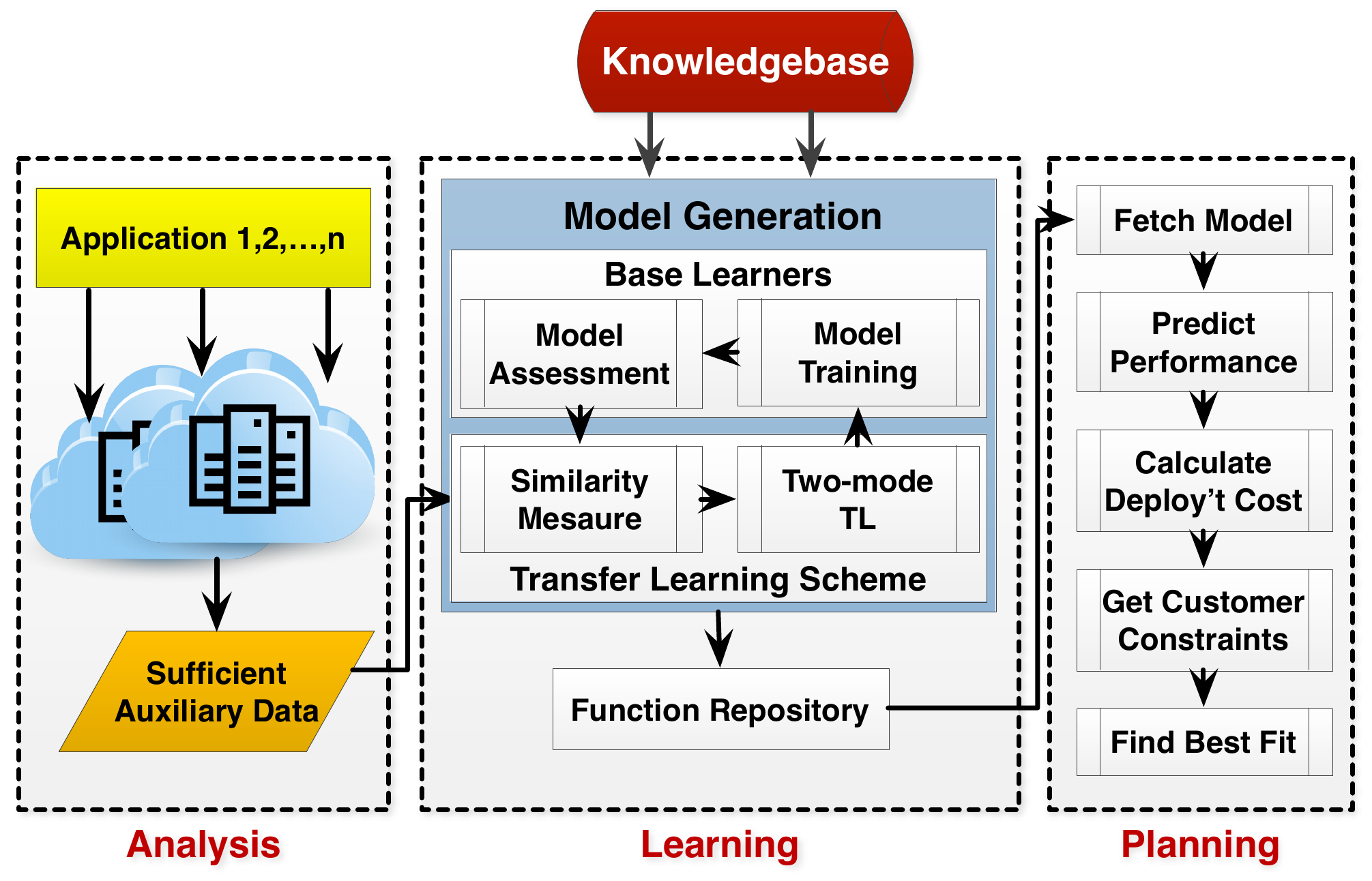}
\caption{The architecture of \thename, a novel TL-assisted cloud DSS.}
\label{fig:DSS-TL detailed}
\end{figure}

The main components of the overall system architecture, shown in \fig{fig:DSS-TL detailed}, are a \textbf{Knowledgebase} and three phases:  \textbf{Analysis}, \textbf{Learning} and \textbf{Planning}. 
The \textbf{Knowledgebase} is comprised of learning methods, prediction models, model parameters along with the data sets the models are derived from. 
The prediction models present in the Knowledgebase are generated by following the traditional ML process (\ie train and test using cross-validation) (\S\ref{sec:base models}). 
The key idea of \thename is to efficiently generate a prediction model for a given application and the process starts with a collection of auxiliary data (\S\ref{sec:auxiliary data}).
The \textbf{Analysis Phase} is responsible for providing auxiliary data to be used by the Learning phase. The auxiliary data is collected by running the application on a representative set of virtual machines and profiling different matrices related to the deployment and performance of the application. 
The \textbf{Learning Phase} is the core of the architecture and is composed of four sub-modules, each performing their own unique tasks to support model generation for a target domain (\ie application and/or provider). 
The model, however, is not generated from scratch by following the long steps of model fitting, but rather it is generated using \thename which makes use of existing learned knowledge fetched from the Knowledgebase. 
\textit{Similarity Measure} helps in identifying the similarity of a new domain (target) with existing ones (source). The auxiliary data is then used by the similarity measure (\S\ref{sec:similarity measure}) to look for a similar application in the Knowledgebase. 
Based on this similarity, \thename transfers existing knowledge using \textbf{two-mode TL scheme}, which is designed to satisfy the goal of enhancing learning efficiency and generating a prediction model for a new application. The `two mode' part of the naming refers to the two possibilities of transferring learned knowledge across domains: \textit{Transfer-All} and \textit{Transfer-Model}. These modes present different means of transferring significant knowledge through the transfer of instance knowledge, feature space knowledge, and parameter knowledge (\S\ref{sec:2TL}).

Together with the similarity outcome and a selected base-learner (\S\ref{sec:base learners}), one of the knowledge transfer modes gets activated. The activated mode feeds in the auxiliary data as well as learned data to the base-learner which then outputs a prediction model.
\textit{Model Training} uses training data sets which may be composed of source and target domain data depending on the activated mode. The test data set for model assessment is comprised of the target domain's data only. 
\textit{Model assessment} is performed using 10-fold cross-validation and, if the output is not satisfactory, the process re-starts by fetching the next most similar application from the Knowledgebase. 
Finally, the accepted model is saved in the Function Repository to be used by the \textbf{Planning Phase} to predict future application performance. Deployment costs are generated from the predicted performance results that are then used to find the best deployment match in accordance with customer-specified constraints.

The remainder of this section details the functionality of \thename modules covering base learners, auxiliary data, and similarity measurement.

\subsubsection{Base learners}
\label{sec:base learners}
Base learners are the ML methods used to derive the initial prediction models for a source task. For \thename, Multivariate Polynomial Regression (MPR), Support Vector Regression (SVR) and Random Forests (RF) are used as base learners.
MPR has the ability to statistically infer and understand the relationship of response and predictor variables related to a data set. Hence, a behavioural relationship for an application can be generated with varying deployment setups.
SVR attempts to obtain a function as flat as possible having an $\epsilon$-deviation from the training output~\cite{Smola2004SVR}.
SVR relies on a set of parameters such as $\epsilon$: the sensitivity to the deviation, the cost function and range of kernels, allowing it to optimize the objective function as well as the parameters of the regressive function. This allows \thename to optimally adjust the hyper-parameter when fed with the different distributional training sets.  
RF ~\cite{Khaled2014RF} is based on an ensemble learning approach making use of Bagging sampling and the random selection of features. The ensemble nature of RF enhances its ability to produce better prediction results. 
In RF, multiple decision trees (ensemble models) are generated using the randomly drawn samples. For unseen data, each decision tree takes part in the prediction process and the final prediction result is the label selected by a majority of decision trees. RF is good in accuracy and gives a better understanding of the variable importance and their correlations. Moreover, this method is considered robust to outliers as compared to other methods.

\subsubsection{Base models}
\label{sec:base models}
\begin{figure}[th]
\centering
\includegraphics[width=0.6\columnwidth,clip,trim=1cm 6.2cm 1cm 1.4cm]{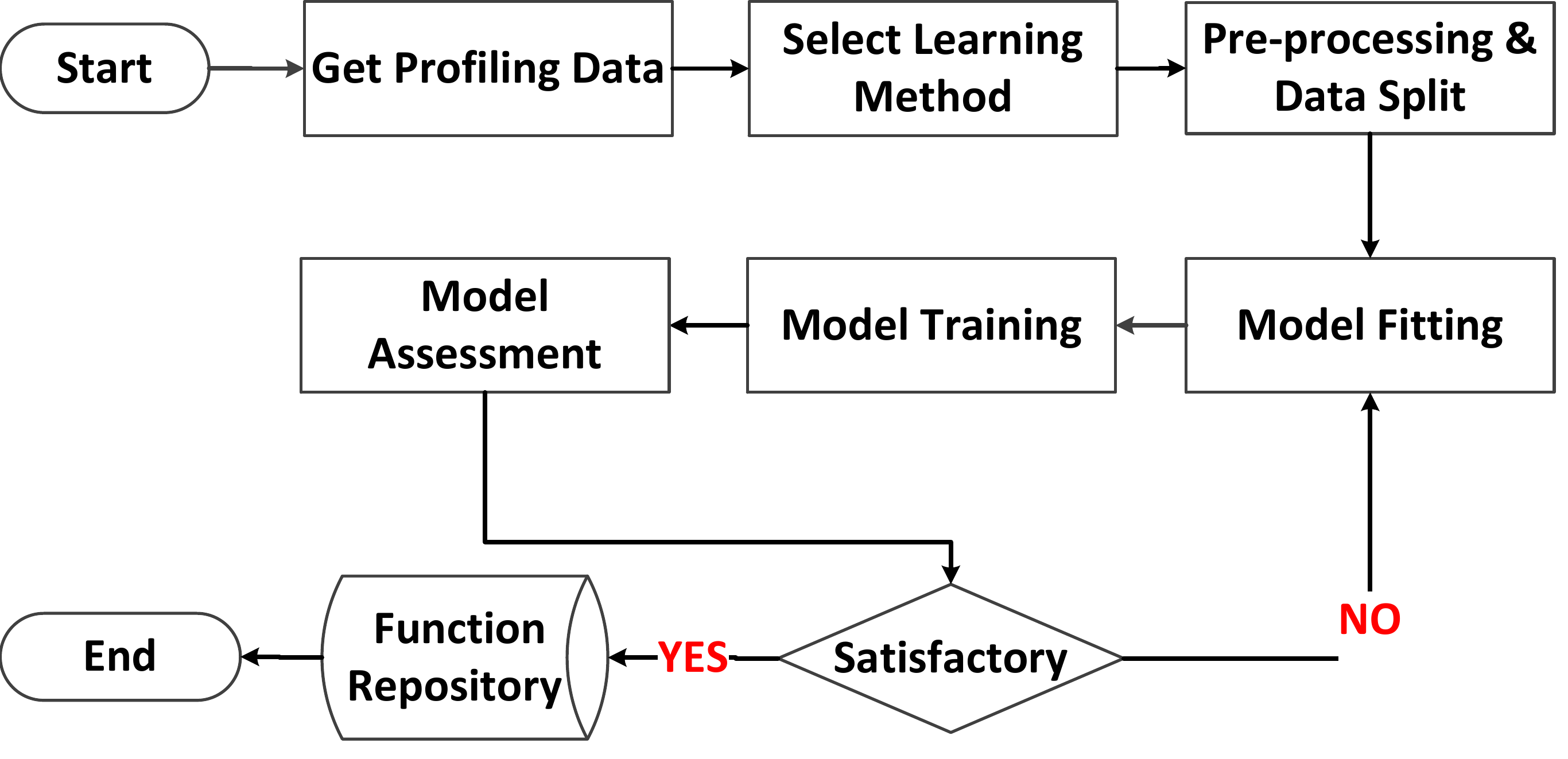}
\caption{The model generation process.}
\label{fig:model_gen}
\end{figure}

Base models are generated by following what is now deemed a traditional ML process of learning a model starting from data collection as detailed in Figure~\ref{fig:model_gen}. This model generation is supported by Daleel's framework~\cite{daleel}, designed to investigate the role of ML to support application-driven decision making around the selection of instance types in a given cloud provider. 

The collected data passes through the pre-processing stage to fulfill initial assumptions about the quality of the data. 
Pre-processed data is then split into training and test sets and fed into an iterative process of fitting a model. The model fitting process is responsible for exploring different distributions to investigate the true relationship of a response variable with the predictors (features). The fitted model is then trained on the training data set using 10-fold cross-validation to capture the maximum variability of data and later passed on to the model assessment process to be assessed for the prediction accuracy. Further to that, the models offering satisfactory prediction capability are then stored in the function repository and, in case of unsatisfactory results, starts updating the model. Base learners are the learning methods involved in the model generation process. 
The significant predictors used are vCPU, RAM, Compute-units, Multithreading, Load-in-memory, Application-Type, External-file-requirement, Parallel-execution, File-size, and Execution-day. Out of these predictors, the first three represent VM configuration and rest are part of application and execution details.

\subsubsection{Auxiliary data}
\label{sec:auxiliary data}
Semi-supervised transductive TL, which \thename is based on, requires the presence of auxiliary data for model generation. 
In this context, the term, ``sufficient'' amount of data is introduced to represent the auxiliary data requirement for the target domain to be used in \thename. This is an important parameter for the approach, derived from extensive experimentation. 
A ``sufficient amount'' is identified by observing model convergence according to the change in percentage contribution of training data.

\subsection{Similarity measurement} \label{sec:similarity measure}
A similarity measure is required to identify which of the source domains (source applications) will give the best performance on the target domains (target applications) to learn the target task (prediction model). 
The proposed approach leverages the Kolmogorov-Smirnov (KS) test, a non-parametric test to compare the cumulative distributions of two data sets in order to identify if the two data samples come from the same distribution. 

The \textit{KS} test is considered effective in comparing samples when there is no available knowledge about the common distribution of the source and target domain data. In addition, this method is sensitive to distribution and there is no restriction on the sample size which makes it useful to capture change in performance even with few samples. These properties make \thename useful for different applications and machine configurations considering that often it is difficult to compare deployment settings and resource utilization between the two applications, especially when the VMs belong to different cloud providers and vary in underlying configurations.

\begin{algorithm}[!th]
	\KwIn{(i) Auxiliary data = $S_{aux}$ \\ (ii) Knowledgebase data = $S_{kb_i}$,\\ where $i= 1...$total number of applications} 
	
	\KwOut{ (i) $D_d$, $D_p$,$D_f$: a similarity estimate for $S_{kb_i}$ \\ (ii) Tagged knowledgebase datasets = $S_{kb_i}.tagged$}
	
	\textbf{Initialization\;}
	\begin{itemize}
		\item 	Let $A_{j1}, .. .,A_{jq},.., A_{jk}$ be the value in $S_{aux}$, where $A_{j1}..A_{jq}$ represents application architecture, \\ and $A_{jq}.. A_{jk}$ represents deployment details \\ 
		
		\item Let $B_{l1}, .. .,B_{lq}, .., B_{lm}$ be the value in $S_{kb_i}$, where $B_{l1}..B_{lq}$ represents application architecture, \\ and $B_{lq}.. B_{lm}$ represents deployment details \\ 
	\end{itemize}

	\textbf{Function} SimilarityMeasure ($S_{aux}$, $S_{kb_i}$) \\
	
	\ForEach{$S_{kb_i} \in \{S_{kb_1}, ... , S_{kb_n}\} $}{	\ForEach{$x \in \{A_{j1},  ..., A_{jk}\} \cup \{B_{l1},  ..., A_{lm}\} $}{ \For{$Two.Sample.KS.Test (x_a, x_b)$}{Compute $p$-value $\rightarrow$ $D_p$ \\  Compute $D^\prime$-value  $\rightarrow$ $D_d$ \\ \eIf{$p$-value ($D_p$) \textgreater 0.05}{$x.mark=``SAME"$ \\$x_p.mark=``SAME"$ \\ $D_f=``NO" $}{\eIf{$D^\prime$ ($D_d$)  \textless  $0.5$}{$x.mark=``SAME"$}{$x.mark=``DIFFERENT"$ \\ $D_f=``YES" $}}}  } $aggregate.x.mark$ for $\{A_{j1}..A_{jq}\}$ and $\{A_{jq}.. A_{jk}\}$ \\ \eIf{value of $aggregate.x.mark$ is ``SAME" for all $\{A_{j1}..A_{jq}\}$ }{$S_{kb_i}.tagged=``SIMILAR"$}{\eIf{ value of $aggregate.x.mark$ is ``SAME" for \textgreater 1/ $\{A_{j1}..A_{jq}\}$}{$S_{kb_i}.tagged=``PARTLY-SIMILAR"$} {$S_{kb_i}.tagged=``NOT-SIMILAR"$}} }
	\caption{Identification of source domain similarity}
	\label{alg:similarity-measure}
\end{algorithm}

Algorithm \ref{alg:similarity-measure} compares the probability distribution of the target application's profiling data ($S_{aux}$) with the existing applications ($S_{kbi}$) and identifies similar distributional application(s) ($S_{kb_i}.tagged$) from the knowledgebase. The application(s) are tagged according to the calculated values for distributional difference ($D_d$) and feature difference ($D_f$).
The \textit{KS}/similarity test is applied on vector inputs ($A_{j1},..., A_{jk}$ \& $B_{l1},.., B_{lm}$ ) from both source and target domain, and each vector input represents a single feature. 
This test outputs a p-value ($D_p$) and a D-value ($D_d$). Here p-value  quantifies the probability of two samples populated from the same or a different distribution and D-value represents the difference of empirical distribution functions of two samples.

If the p-value rejects the similarity hypothesis, then the D value is evaluated to get an idea about the probability of similarity. A D-value in the range of 0.0--0.5 is considered as a measure of corresponding sample similarity from 100\% to 50\%. The aggregated p-values and D-values assist in deciding the similarity. 

Table~\ref{tab:kstest result} presents the outcome of a similarity analysis based on three real world applications. The first 7 features in the table represent cloud and deployment related information and the remaining features explain application architecture. The three applications are named anonymously as A, B and C. The D-value and p-value is calculated for each feature vector taken from the source and target applications as explained in the \texttt{SimilarityMeasure} function. The similar features are marked as "*". An application with a high number of similar features will get a higher ranking in terms of similarity. The resulting similarity is tagged as one of three categories: 1) Similar, 2) Partly-Similar, or 3) Not-Similar.

The aggregated similar features for A and B is higher than A and C so, for application A being a target application, B would be the first choice to be used as a source of knowledge. Moreover, applications A and B have similar architectures and are tagged as \textit{Partly-Similar} to each other, however application C is tagged as \textit{Not-Similar}. For application C both the applications A and B are ranked the same to be used as the source, however both applications have almost no similarity at application architecture level and hence will be tagged as \textit{Not-Similar} to each other. 

\begin{table}[tb]
\scriptsize
\centering
\caption{Pairwise similarity analysis among three applications: A, B, and C. Similarity is indicated by p- and D-values, which are calculated using the KS method for cloud, deployment and application specific features.}
\begin{tabular}{caaccaa}
\toprule
\rowcolor{Gray!50} &\multicolumn{2}{a}{\textbf{A \& B}} &\multicolumn{2}{c}{\textbf{A \& C}} &\multicolumn{2}{a}{\textbf{C \& B}}\\
\rowcolor{Gray!50} \textbf{Feature} & \textbf{D-value} & \textbf{p-value} & \textbf{D-value} & \textbf{p-value} & \textbf{D-value} & \textbf{p-value}\\
\midrule
vmtype & 0.4082 * & 2.20E-16 & 0.3986 * & 2.20E-16 & 0.662 & 2.56E-07\\
\hline
vcpu & 0.4063 * & 2.20E-16 & 0.431 * & 2.20E-16 & 0.0662 * & 2.56E-07\\
\hline
ecu &  0.4063 * & 2.20E-16 & 0.431 * & 2.20E-16 & 0.0662 * & 2.56E-07 \\
\hline
ram & 0.4063 * & 2.20E-16 & 0.431 * & 2.20E-16 & 0.0662 * & 2.56E-07 \\
\hline
day & 0.0804 * & 2.32E-16 & 0.0926 * & 2.20E-16 & 0.173 * & 2.56E-07 \\
\hline
subm. time & 0.4063 * & 2.20E-16 & 0.431 * & 2.20E-16 & 0.0662 * & 2.56E-07\\
\hline
exec. time & 1 & 2.20E-16 & 1 & 2.20E-16 & 0.633 & 2.20E-16 \\
\hline
app. type & 0.4063 * & 2.20E-16 & 0.431 * & 2.20E-16 & 0.0662 * & 2.56E-07\\
\hline
multi-threading& 0.4063 * & 2.20E-16 & 0.431 * & 2.20E-16 & 0.0662 * & 2.56E-07\\
\hline
external files &  0 * & 1 ** & 1  & 2.20E-16 & 1  & 2.20E-16\\
\hline
loaded in mem. &  1 & 2.20E-16   & 1  & 2.20E-16 & 0 * & 1 **\\
\hline
parallel &  0 * & 1 ** & 1  & 2.20E-16 & 1  & 2.20E-16\\
\hline
file size & 1  & 2.20E-16  & 1  & 2.20E-16 & 1  & 2.20E-16\\
\hline
\rowcolor{Gray!50}\textbf{Result}&\multicolumn{2}{a}{\textbf{Partly similar}} &\multicolumn{2}{c}{\textbf{Not similar}} &\multicolumn{2}{a}{\textbf{Not similar}}\\
\bottomrule
\end{tabular}
\label{tab:kstest result}
\end{table}

\subsection{The Two-mode TL scheme}
\label{sec:2TL}
This section explains the core TL algorithm, derived from extensive experimental analysis on public cloud providers and applications of varying requirements. 
The two modes and the respective knowledge transfer approaches are discussed here:

\begin{enumerate}
\item {\textbf{Transfer-All}} includes three approaches to transfer knowledge from the source to the target domain. 
\begin{enumerate}
\item Transferring feature representation knowledge 
\item Transferring instance knowledge 
\item Transferring parameter knowledge 
\end{enumerate}

\item {\textbf{Transfer-Model}} includes two approaches to transfer knowledge from source to target domain.
\begin{enumerate}
\item Transferring feature representation knowledge
\item Transferring parameter knowledge 
\end{enumerate}
\end{enumerate}

\begin{algorithm}[!t] \label{alg:two-mode-scheme}
\KwIn{(i) Auxiliary data = $S_{aux}$ \\ (ii-a) Tagged knowledgebase datasets = $S_{kb_i}.tagged$ \\ (ii-b) $X_{sig}$ of $S_{kb_i}.tagged$\\ (ii-c)  Learning Method = $M_{kb_i}$, where $M$ is SVR, MPR, RF \\ (ii-d) PredictionFunction = $f(S_{kb_i})$ \\  } 
\KwOut{ (i) PredictionFunction = $f(S_{aux})$}
\textbf{Function} \thename($S_{aux}$, $S_{kb_i}.tagged$, $M_{kb_i}$) \\

\ForEach{$S_{kb_i}.tagged$}{\eIf{$S_{kb_i}.tagged$ == ``SIMILAR" $\|$ $S_{kb_i}.tagged$ == ``PARTLY-SIMILAR" }{Set $D_d$ = FALSE\\ \eIf{$M_{kb_i}$ == SVR $\|$ RF}{Set BaseLearner=$M_{kb_i}$ \\ Call TrasnferAll($S_{aux}$, $S_{kb_i}.tagged$, $X_{sig}$, BaseLearner, $D_d$, $f(S_{kb_i})$)}{\If{$M_{kb_i}$ == MPR}{Set BaseLearner=$M_{kb_i}$ \\ Call  TrasnferModel($X_{sig}$, $S_{aux}$, $f(S_{kb_i})$, BaseLearner, $D_d$)}{}}}{\If{$S_{kb_i}.tagged$ == ``NOT-SIMILAR"}{ Set $D_d$ = TRUE\\ \eIf{$M_{kb_i}$ == SVR $\|$ RF}{Set BaseLearner=$M_{kb_i}$ \\ Call TrasnferModel($X_{sig}$, $S_{aux}$, $f(S_{kb_i})$, BaseLearner, $D_d$)}{\If{$M_{kb_i}$ == MPR}{Set BaseLearner=$M_{kb_i}$ \\ Call TrasnferModel($X_{sig}$, $S_{aux}$, $f(S_{kb_i})$, BaseLearner, $D_d$)}}
}}}

\caption{Two-mode TL Scheme}
\end{algorithm}

\thename works by activating one of its modes based on the inputs from similarity measurement and a base learner. If the similarity outcome select a "Similar" or "Partly-Similar" application as a source of knowledge and if the learning method is SVR or RF then \textit{Transfer-All} is activated and the respective knowledge is transferred from the source application to the target application. On the other hand, if there is a ``Not-similar" application and the learning method is still SVR or RF, \textit{Transfer-Model} gets activated. In case of the MPR learning method, the only possibility for activation mode is \textit{Transfer-Model}. We now explain each of the above approaches involved in the designed scheme. 

\vspace{0.25em}\noindent\textbf{\textit{(a) Transferring feature representation knowledge }}

\noindent The feature space represents specific properties related to application architecture, deployment configurations and execution details. 
If both the source and target domains have some similarity at the application or deployment level, the higher the chances are for generating a reasonable target-learning model using the same features as for the source model. Therefore, `significant' features are transferred from the source domain to the target domain. 
If the feature space differs in both source and target domains due to the varying standards of IaaS offerings, mapping of similar features is required. The mapping is done manually using the shared knowledge of both domains. 

\vspace{0.25em}\noindent\textbf{\textit{(b) Transferring instance knowledge }}

\noindent The instance knowledge represents a sample set comprised of the selected feature space. 
If the source and target domains have high similarity then the instance knowledge transfer can positively contribute for the model generation of the target domain, especially when there are few sample sets available for the target domain. However, the best practice is to transfer instances in an incremental manner in order to avoid any influential effect from the source data.

\vspace{0.25em}\noindent\textbf{\textit{(c) Transferring parameter knowledge}}

\noindent Parameter knowledge details the mathematical formulation of the estimation or prediction function. The parameter knowledge can be shared with the target domain in accordance with the selected base learner. If the base learner is SVR then the kernel function, tuning parameters' learning rates, the learning-cost function and model constants are transferred. For MPR, the polynomial order along with coefficient values and any interaction terms are transferred. In the case of RF, information about the number of decision trees to be generated and a number of variables to be tried at each step are shared with the target domain.

\section{Experimental Evaluation}
This section describes the overall strategy to assess \thename and offers a comprehensive evaluation of the proposed scheme. 
\thename is evaluated using two public cloud providers with three representative real-world applications. These applications differ in their underlying architectures and also their memory and CPU utilization. 

\subsection{Evaluation strategy}
\begin{figure}[bt]
\centering
\includegraphics[width=0.4\columnwidth]{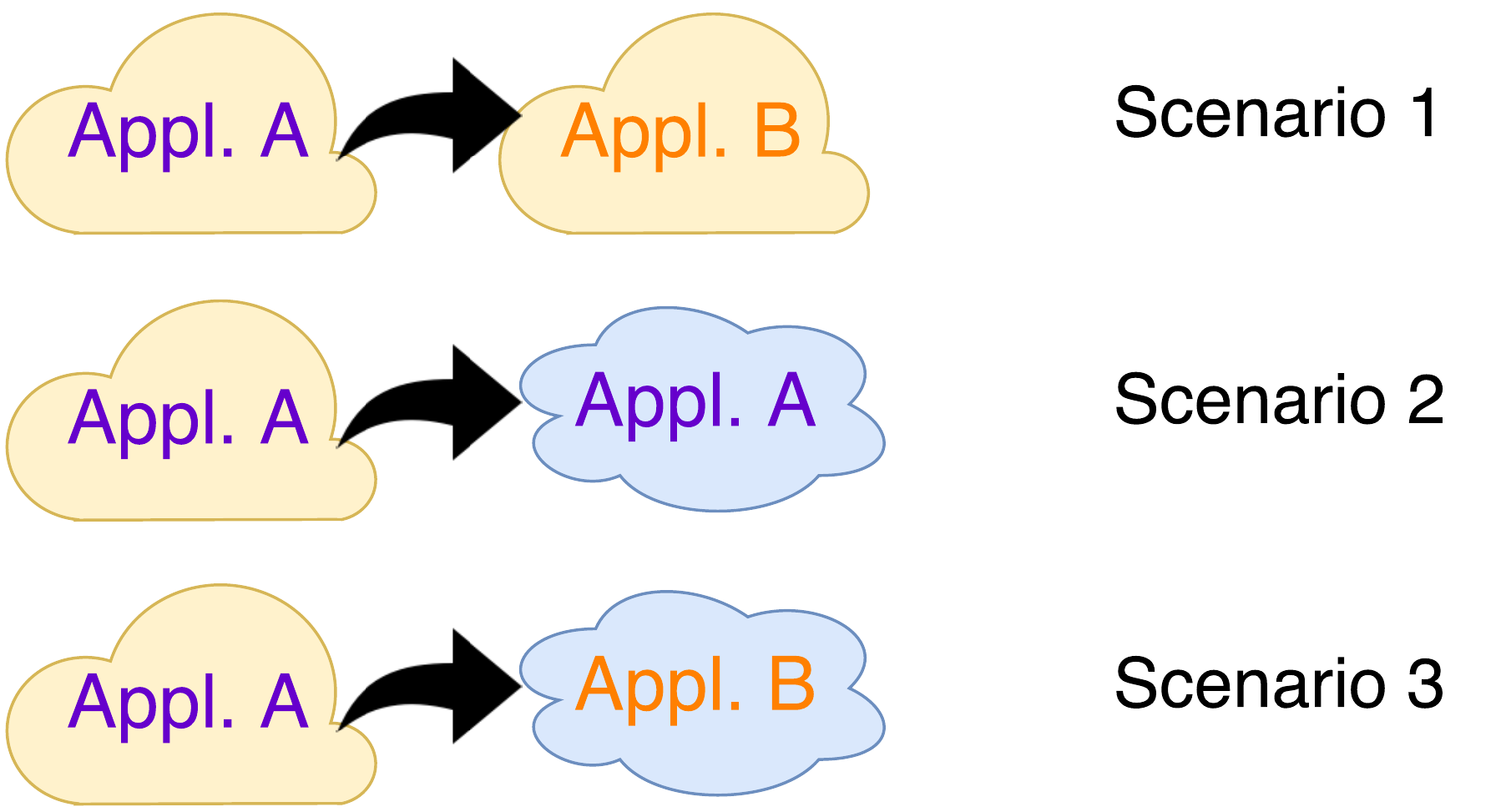}
\caption{A high-level description of the three evaluation scenarios.}
\label{fig:eval-sc}
\end{figure}

Our evaluation criteria are as follows: 
\begin{enumerate}
	\item \textbf{Accuracy of base models}, \ie the traditional ML approach, in terms of prediction performance using overall data and using daily subsets thereof.
	\item \textbf{Feasibility of our TL approach} under 3 distinct different evaluation scenarios: transferring to a new application, to a new provider, and to both a new application and provider, respectively (\fig{fig:eval-sc}). 
	\item \textbf{Time and cost effectiveness} of the proposed scheme in terms of incurred cloud costs and model generation time. 
\end{enumerate}

All application parameters and input are kept constant between application runs to reduce the dimensionality of the experiments. In order to collect enough data for learning purposes, we repeat a workload with a 10 minute delay between each pair of runs. This is spread over different times of the day and different days of the week to capture temporal variance. The Linux tools \texttt{vmstat}, \texttt{glances} and \texttt{sysstat} are used to continuously monitor resource utilization.

Different evaluation metrics were computed to assess the model accuracy, namely $R^2$, Residual Standard Error (RSE), Mean Squared Error (MSE) and Percentage Relative Root Mean Squared Error (\%RRMSE). 
The metrics presented in this paper are MSE and \%RRMSE. MSE is a standard measure of prediction error and we use it to assess individual model performance. \%RRMSE provides a fair comparison of prediction accuracy between different learning models and data sets with variable scales by calculating a scale-independent model error relative to the actual value. As such, we use it to evaluate the performance of different models for different data sets. 
\%RRMSE is expressed as:
\begin{equation}\label{eq:RMSE}
\%RRMSE = \sqrt{\frac{1}{N}\sum_{t=1}^{N}{\frac{(y^i - y)^2} {y}}} * 100
\end{equation}
where $N$ is the total number of samples, and $y^i$ and $y$ are the predicted and actual value of application execution time, respectively. The lower the \%RRMSE value is the better the model performance.

\subsection{Applications}
\label{sec:apps}
We selected 3 open-source applications of different architectures and requirements of CPU and memory usage.

\begin{enumerate}
\item \textbf{VARD} is 
a tool designed to detect and tag spelling variations in historical texts, particularly in Early Modern English \cite{baron2008Vard2}. The output is aimed at improving the accuracy of other corpus analysis solutions. Vard is a single threaded Java application that is highly memory intensive. It holds in memory a representation of the full text, as well as various dictionaries that are used for normalizing spelling variations.
Vard is considered a pre-processor tool to other corpus linguistic tools such as keyword analysis, semantic tagging and annotations \etc

\item \textbf{smallpt} 
is a global illumination renderer written as a multi-threaded OpenMP application and, as such, is considered CPU intensive. The application performs unbiased Monte Carlo path tracing to calculate the amount of light reaching different points in a given scene, offering features such as anti-aliasing, soft shadows, and ray-sphere intersection. 
For this research we selected a box scene that is constructed out of nine very large overlapping spheres.

\item \textbf{ItemRecommender}
uses a collaborative filtering technique in order to recommend similar items, as identified using log-likelihood similarity. ItemRecommender is written as a single-threaded Java based program, and uses the MovieLens\footnote{https://grouplens.org/datasets/movielens} dataset collected and maintained by GroupLens. The dataset is comprised of 10 million ratings of 10,000 movies by 72,000 users. This is used to to recommend movies to a user based on the preferences of other users. The program takes every other person who has reviewed at least 5 movies in common with the user and calculates the Pearson correlation between these 2 users. Accordingly, weights are calculated and converted to numerical values to be assigned to the user's scale and finally the recommendations are listed after sorting.
\end{enumerate}

\subsection{Cloud infrastructure}
For our target cloud providers we use Amazon EC2 and Google GCE, two of the major providers in the IaaS market. From each, we selected a subset of their on-demand (\ie pay-as-you-go) Linux instance types deemed suitable to run the aforementioned applications. These instance types are summarized in Tables \ref{tab:instance-types-ec2}-\ref{tab:instance-types-gce}.

\begin{table}[!tb]
\footnotesize
\centering
\caption{Computational specifications of EC2 instances.}
\begin{tabular}{ccccccc}
\toprule
\textbf{Series} & \textbf{Instance} & \textbf{vCPU} & \textbf{ECU} & \textbf{RAM} & \textbf{Storage} & \textbf{Price}\\
&\textbf{Type}&&&(GiB)&(GB)& (\$/h)\\
\midrule
T2 (General & \texttt{t2.small} & 1 & Var. & 2 & 20 & 0.026\\
\cline{2-7}
Purpose) & \texttt{t2.medium} & 2 & Var. & 4 & 20 & 0.052\\
\hline
M3 & \texttt{m3.medium} & 1 & 3 & 3.75 & 4(S) & 0.070\\
\cline{2-7}
(General & \texttt{m3.large} & 2 & 6.5 & 7.5 & 32(S) & 0.140\\
\cline{2-7}
Purpose) & \texttt{m3.xlarge} & 4 & 13 & 15 & 32(S) & 0.280\\
\hline
C4 & \texttt{c4.large} & 2 & 8 & 3.75 & 20 & 0.116\\
\cline{2-7}
(Compute & \texttt{c4.xlarge} & 4 & 16 & 7.5 & 20 & 0.232\\
\cline{2-7}
Optimised) & \texttt{c3.xlarge} & 4 & 14 & 7.5 & 32(S) & 0.239\\
\bottomrule
\end{tabular}
\label{tab:instance-types-ec2}
\end{table}

\begin{table}[!tb]
\footnotesize
\centering
\caption{Computational specifications of GCE instances.}
\begin{tabular}{ccccccc}
\toprule
\textbf{Series} & \textbf{Instance} & \textbf{vCPU} & \textbf{GCEU} & \textbf{RAM} & \textbf{Storage} & \textbf{Price}\\
&\textbf{Type}&&&(GB)&(GB)& (\$/h)\\
\midrule
Standard & \texttt{n1-standard-1} & 1 & 2.75 & 3.75 & 16 & 0.036\\
\cline{2-7}
Type & \texttt{n1-standard-2} & 2 & 5.5 & 7.5 & 16 & 0.071\\
\cline{2-7}
& \texttt{n1-standard-4} & 4 & 11 & 15 & 16 & 0.142\\
\hline
High Mem. & \texttt{n1-highmem-2} & 2 & 5.5 & 13 & 16 & 0.106\\
\hline
High & \texttt{n1-highcpu-2} & 2 & 5.5 & 1.8 & 16 & 0.056\\
\cline{2-7}
CPU & \texttt{n1-highcpu-4} & 4 & 11 & 3.6 & 16 & 0.118\\
\cline{2-7}
& \texttt{n1-highcpu-8} & 8 & 2.2 & 7.2 & 16 & 0.215\\
\bottomrule
\end{tabular}
\label{tab:instance-types-gce}
\end{table}

Both Amazon EC2 and GCE provide a differentiated series of instance types, catering to different application needs (\eg compute-intensive, memory intensive, I/O-intensive, and so on). Each series contains a number of instance types offering different setups of computational resources. From EC2 we targeted the General Purpose T2 and M3 series. 
In addition, the Compute Optimized series is also selected with three representative instance types. 
Similarly, from GCE we targeted the Standard Type series instances.
In addition, we selected the High CPU series 
as well as the High Memory series 
in order to evaluate varying combinations of resource capacities over a relatively wide price range. 

All instances used run 64-bit Ubuntu Linux of different capacities as shown in Tables  \ref{tab:instance-types-ec2}--\ref{tab:instance-types-gce}. 
GCE differs from  Amazon-EC2 in various aspects such as the pricing scheme, VM configuration measurement units and compute units. Amazon charges on an hourly basis for VMs; in contrast, Google charges a minimum of 10 minutes per VM. Both providers have non-standard categories to offer the pool of VM's computational power and units.
The Google compute engine unit (GCEU) is an abstraction of compute resources. According to Google, 2.75 GCEUs represent the minimum power of one logical core.
Amazon uses the term `ECU' as a computation unit to express the CPU capabilities of its various compute offerings. 
The capacity unit for measuring the disk size, machine type memory, and network usage are calculated in gigabytes (GB) for each EC2 instance of Amazon. In contrast, GCE uses gibibyte (GiB) as a measuring unit for describing configurations of the VMs.
It is very hard to make a 1:1 comparison with such non-standard and sometimes vague descriptions about the computational units. This creates a difference in terms of the feature space at the domain level. Moreover, some of the information, such as details of parallel workload on VMs, scheduling algorithms and how GCE virtual cores are pinned to physical cores, is not provided by the public cloud providers. So, users of IaaS cannot perceive any collocation or interference effect on their running application. 
The proposed approach transcends these problems and deals with such differences at the feature space level by mapping of similar features.

\section{Results}

\subsection{Overall accuracy of base models}

\begin{figure}[b]
\centering
\includegraphics[width=0.5\columnwidth,clip,trim=0 1cm 0 0]{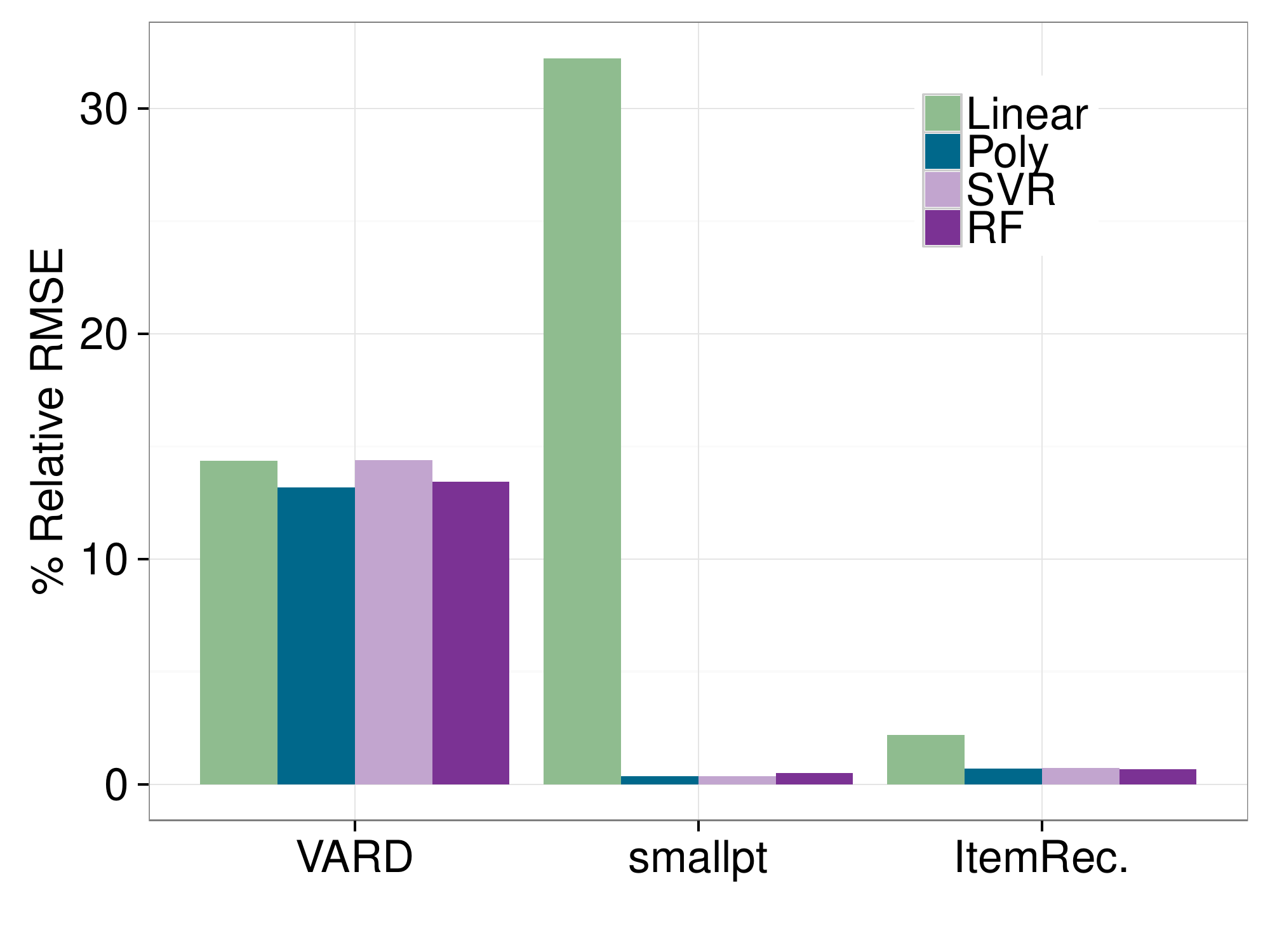}
\caption{\%RRMSE comparison of Linear, Polynomial, SVR and RF based models generated for VARD, smallpt \& ItemReccomender.}
\label{fig:base-model-asessment}
\end{figure}

\begin{figure*}[th]
\centering
\begin{tabularx}{1\textwidth}{@{}ccc@{}}

\includegraphics[width=0.32\textwidth, trim=0cm 0.2cm 0.2cm 0.25cm, clip]{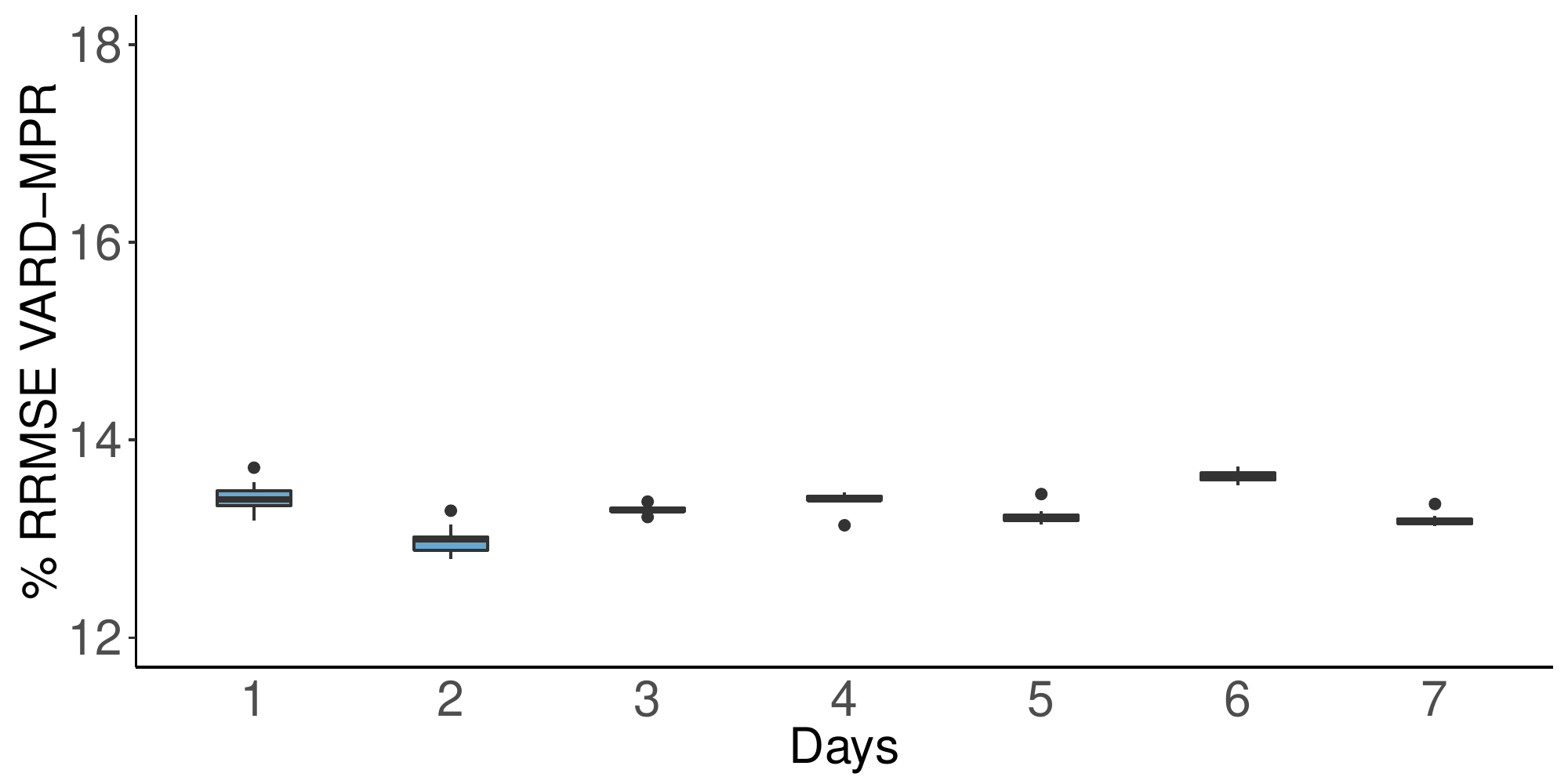}
&
\includegraphics[width=0.32\textwidth, trim=0cm 0.2cm 0.2cm 0.25cm, clip]{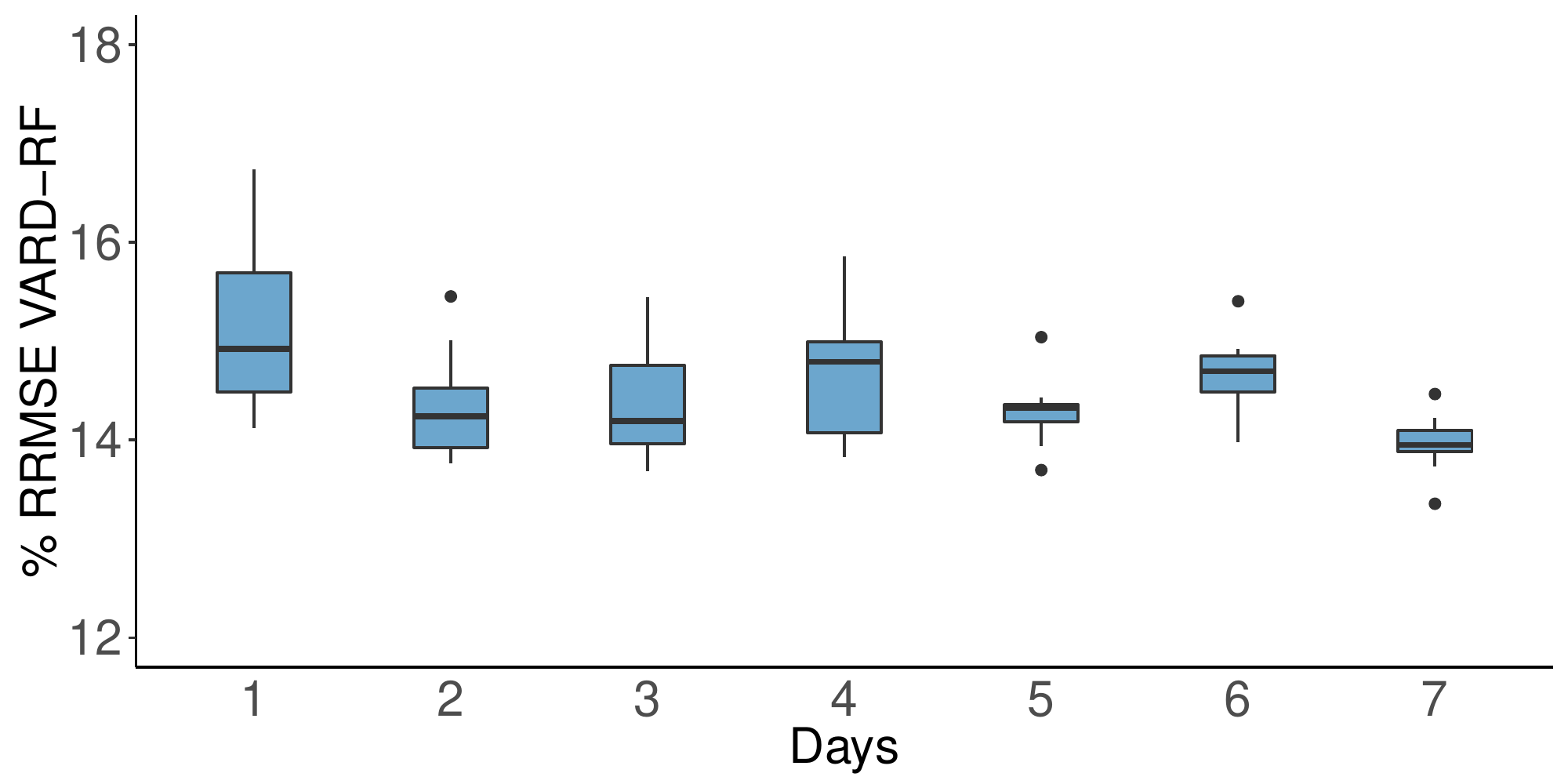}
&
\includegraphics[width=0.32\textwidth, trim=0cm 0.2cm 0.2cm 0.25cm, clip]{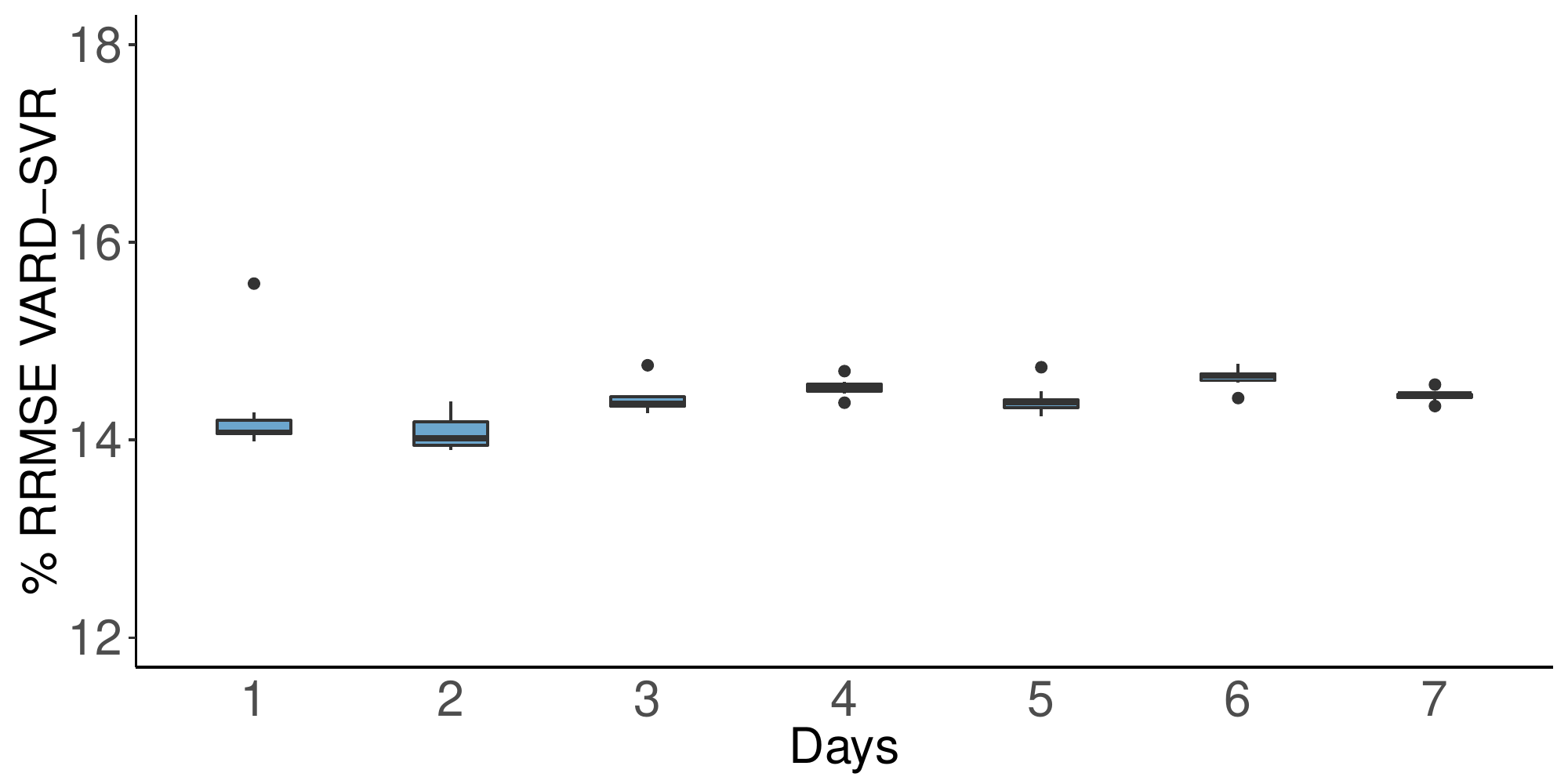}
\end{tabularx}

\caption{Model error (\%RRMSE) to assess per day prediction accuracy of three base models of VARD application over the span of 7 days. The models displayed from left to right are: MPR, RF and SVR.}
\label{fig:per day accuracy Vard}
\end{figure*}

\begin{figure*}[th]
\centering
\begin{tabularx}{1\textwidth}{@{}ccc@{}}

\includegraphics[width=0.32\textwidth, trim=0cm 0.2cm 0.2cm 0.25cm, clip]{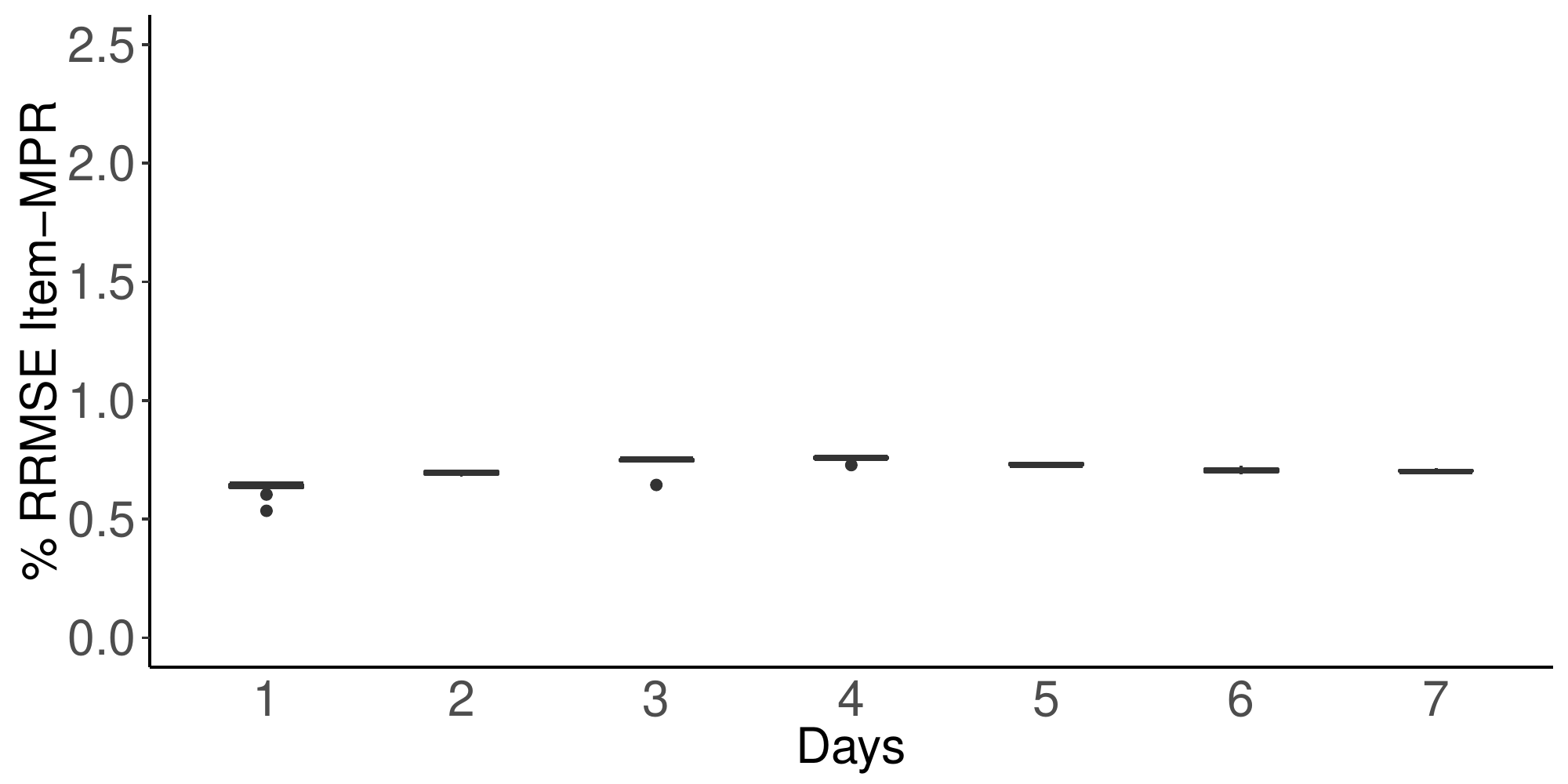}
&
\includegraphics[width=0.32\textwidth, trim=0cm 0.2cm 0.2cm 0.25cm, clip]{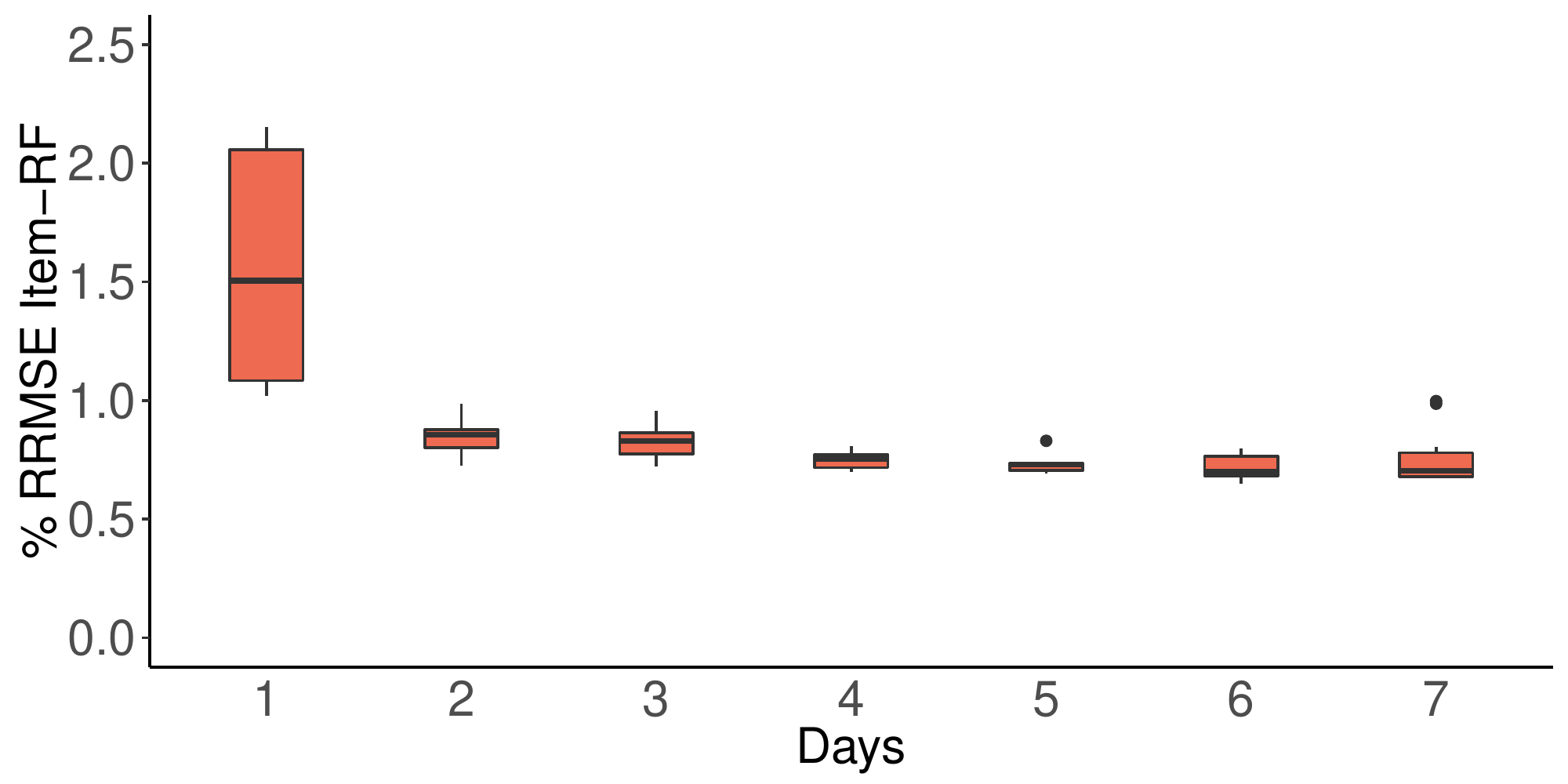}
&
\includegraphics[width=0.32\textwidth, trim=0cm 0.2cm 0.2cm 0.25cm, clip]{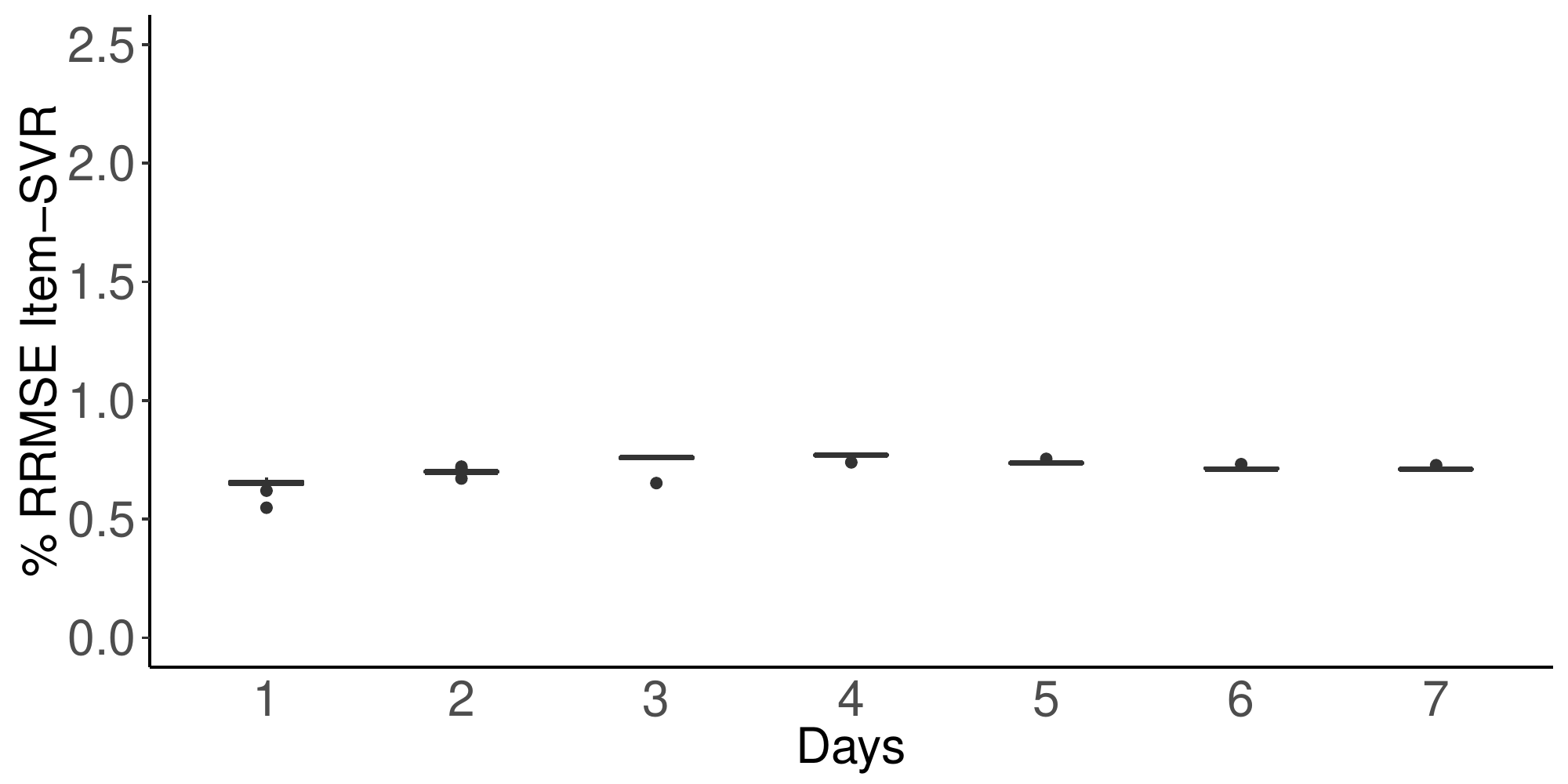}
\end{tabularx}

\caption{Model error (\%RRMSE) to assess per day prediction accuracy of three base models of Item Recommender application over the span of 7 days. The models displayed from left to right are: MPR, RF and SVR.}
\label{fig:per day accuracy item}
\end{figure*}

\begin{figure*}[th]
\centering
\begin{tabularx}{1\textwidth}{@{}ccc@{}}

\includegraphics[width=0.32\textwidth, trim=0cm 0.2cm 0.2cm 0.25cm, clip]{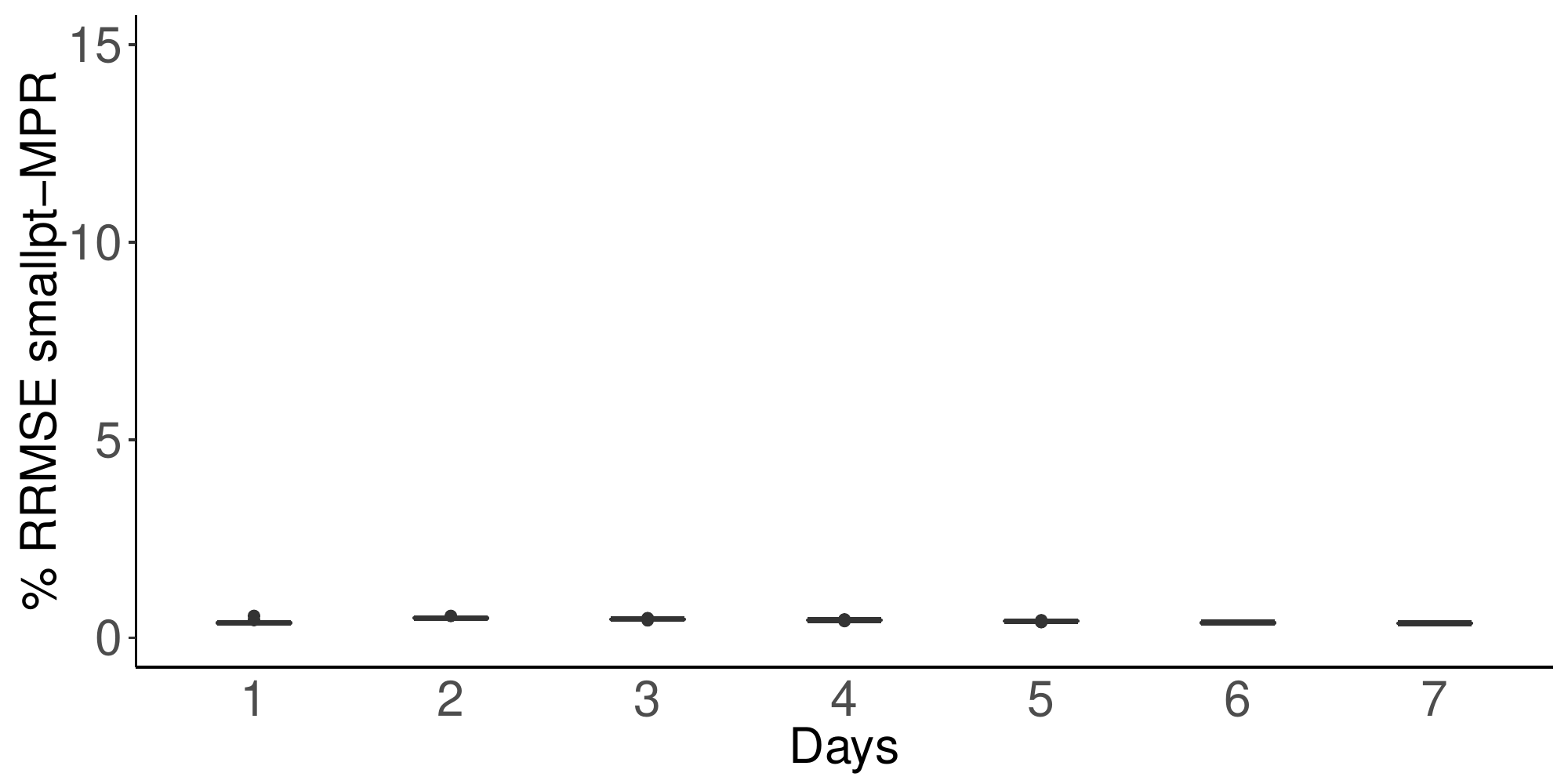}
&
\includegraphics[width=0.32\textwidth, trim=0cm 0.2cm 0.2cm 0.25cm, clip]{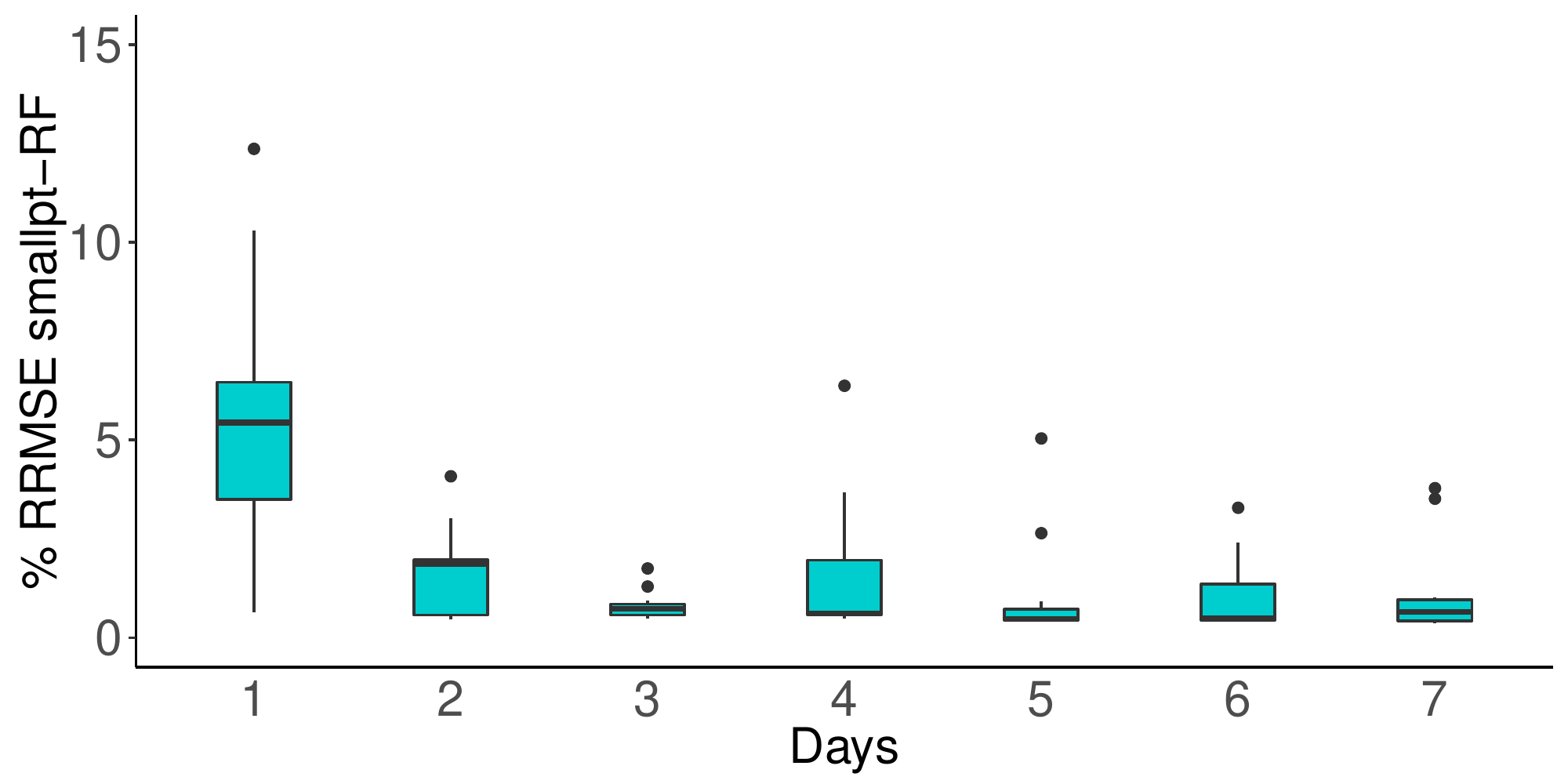}
&
\includegraphics[width=0.32\textwidth, trim=0cm 0.2cm 0.2cm 0.25cm, clip]{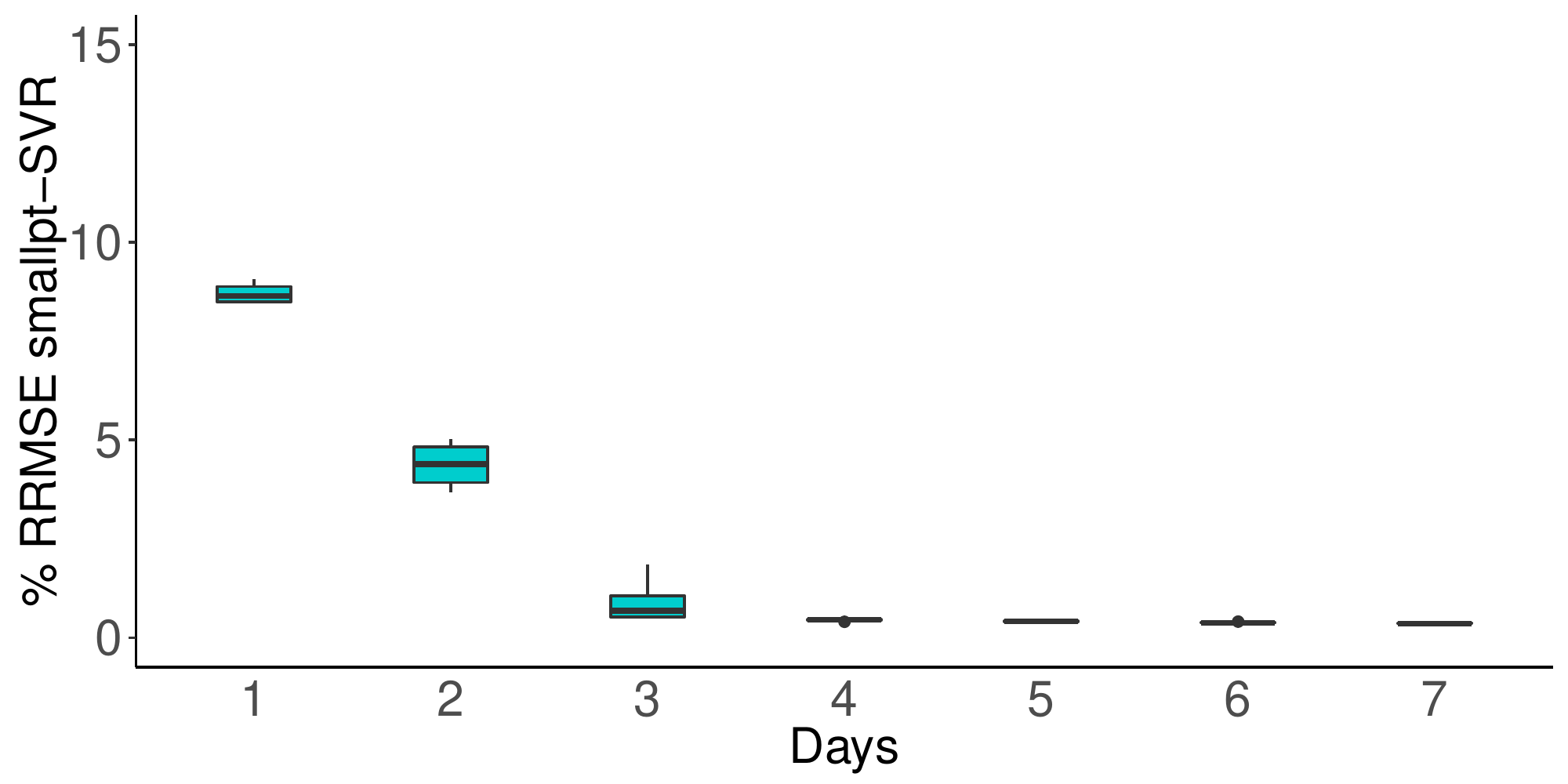}
\end{tabularx}

\caption{Model error (\%RRMSE) to assess per day prediction accuracy of three base models of smallpt application over the span of 7 days. The models displayed from left to right are: MPR, RF and SVR.}
\label{fig:per day accuracy small}
\end{figure*}

Base models are generated using a traditional ML approach as explained in \sect{sec:base models} and a range of ML methods are involved in this process. The generated models for each application are compared with each other in order to assess the prediction accuracy. \fig{fig:base-model-asessment} displays the \%RRMSE of multiple prediction models for three applications as bar plots. 
For ItemRecomender, the prediction errors of MPR, SVR and RF-based models are of very similar error values, all being quite low (<0.72\%). Similarly, MPR, SVR and RF-based models for smallpt are the ones with the lowest model error across (<0.41\%). 
However, the same prediction pattern can be observed for Vard with values for MPR, SVR and RF-based models of 13.300\%, 14.459\%, and 13.472\%, respectively, a little higher compared to smallpt and ItemRecomender specific models.
The low \%RRMSE values for these generated models indicate considerably reduced prediction error. In contrast, the linear model has a slightly higher error rate for Vard and ItemRecommender and a much higher one for smallpt, indicating that this is a poor model in terms of prediction accuracy.

We now explore the variance of model performance on a daily basis to assess the amount of sufficient data needed for accurate predictions. MPR, SVR, and RF based models are also evaluated to assess per day prediction accuracy for three applications, as shown in Figures \ref{fig:per day accuracy Vard}--\ref{fig:per day accuracy small}. The box plots show the spread of \%RRMSE values (y-axis) for 7 days of data where each entry is a result of 10-fold cross-validation. 

The left plot in \fig{fig:per day accuracy Vard} displays the prediction error of the MPR-based model for Vard with a minor variability at a day level. The error becomes more consistent with the addition of more data samples for each day and the majority of data samples are gathered around the third quartile and the first quartile for day1 and day2, respectively. The variability at distributional level gives an indication of performance variation at the VM level, investigated in more detail below.
The MPR-based model for the item recommender in \fig{fig:per day accuracy item} shows very low error with consistent prediction results for each day where most of the data samples are lying around the median. The left plot in \fig{fig:per day accuracy small} represents the model error of the MPR-based model for smallpt and surprisingly a lowermost error rate is observed with very steady predictions for each day. The low error rates confirm that the model is able to capture maximum data variability with accurate prediction for the majority of data samples. 

On the other hand, RF-based models are displaying a common pattern in all boxplots, as shown in the middle plots of Figures \ref{fig:per day accuracy Vard}--\ref{fig:per day accuracy small}. A big drop in the model error is observed from day 1 to day 2 and minor drops from day 2 onwards. An interesting observation is the variability pattern for RF-Vard which is similar to MPR-Vard and providing further evidence of our assumption about significant performance fluctuation. For the RF-based model for ItemRecommender, the middle plot in \fig{fig:per day accuracy item}, we observe a consistent decline for \%RRMSE with a narrow spread at quartile ranges and the median values are nearly aligned with the MPR-based model. The middle plot in \fig{fig:per day accuracy small} shows a contrasting behaviour (RF-smallpt) compared to MPR-smallpt. The error values are spread across different quartiles on a wide range for day1 and \%RRMSE keeps reducing until day 3 and then the median is consistently at the same level for the rest of the days.

Similar to MPR-Vard and RF-Vard, SVR-Vard has the same variable pattern for model error, as shown on the right in \fig{fig:per day accuracy Vard}. Contrary to RF-Vard, the error spread in SVR-Vard is narrow and the majority of data samples are gathered around the median. The model error for the SVR-based model for ItemRecommender (right plot in \fig{fig:per day accuracy item}) is almost the same as for the MPR based model, making it more consistent for prediction accuracy. Finally, the plot for smallpt as shown in \fig{fig:per day accuracy small} displays the highest model error (>9) for day 1 with this gradually decreasing down to less than 1 by day 3 and then this is consistent until day 7. All three base learners are showing similar results with respect to different applications and presenting comparable prediction performance. Overall, MPR is converging to a lower error rate with just a day sample and has less variable prediction performance. SVR also has nearly the same pattern and converges quicker to a lower error value for two of the applications (Vard and ItemRecommender), however, there is not much evidence regarding giving a better performance with more data samples. RF, on the other hand, is more variable in terms of spread and offers better results with more data samples.

\subsection{Feasibility and assessment of the approach}
We now assess the ability of \thename to enhance learning efficacy across domains. We discuss each evaluation scenario (\fig{fig:eval-sc}) in turn.

\subsubsection{Scenario 1: Cross-application} \label{sec:Evaluation 1}
This first evaluation is to ascertain whether \thename is able to satisfy the objective of enhancing the learning efficiency in a cross-application scenario, which could be explained as such: 
\begin{mybox}
A model for application B can be generated to predict its performance on cloud X using the learned knowledge of application A having a model to predict its performance on cloud X.
\end{mybox}

Table \ref{tab:perc-relative-mse-ec2} displays the \%RRMSE of the MPR, SVR and RF-based models for this scenario, where the left most column indicates the source applications with respect to target applications listed in each column. The source applications are used as the source of knowledge and \thename activates the required transfer learning mode according to similarity outcomes. 
Each cell value in the table is the \%RRMSE value for each model, except highlighted cells which correspond to the model performance when \thename is compared to the base model. The column labeled as `smallpt' is referred to as column1 (c1) and so the base model values for smallpt are listed in column1 x row1 (c1r1) and so on.
Here the term base model indicates the model that is generated using traditional machine learning methods (section:\ref{sec:base models}), and training and test data sets are drawn from the same application data for which the model is generated. 

C1 displays the prediction error of models generated for smallpt (target application). Vard and ItemRecommender are considered as source applications and have no similarity at application level resulting in activation of Transfer-Model. This is the case when the model specifics, as well as parameters, are used to train a model for the target application. 
The \%RRMSE displayed in c1r2 and c1r3 are the prediction errors of the models generated by \thename using the knowledge from Vard and ItemRecommender, respectively.
The transferred models are performing the same as the base models (c1r1) and a negligible difference is observed for the model errors. The percentage difference for MPR-based models is 0.002\%-0.008\%, for SVR the range is 0.006\%-0.008\% and for RF the difference is 0.089\%. 

C2 represents the model results for Vard being a target application and smallpt (c2r1) and ItemRecommender (c2r3) as the source, where Vard shows partial similarity with the later source resulting in the use of the Transfer-All mode (section: \ref{sec:2TL}). The observed difference of model error for MPR is 0.008\%-0.015\%, SVR is 0.029\%-0.022\% and RF is from 0\%-0.052\%.

The right most column (c3) follows the same output trend, where ItemRecommender is the target application with smallpt and Vard as a source. The \% difference of RRMSE values (c3r1, c3r2) for MPR, SVR and RF-based models is 0.002\%-0.025\%, 0.001\%-0.002\% and 0.024\%, respectively.

\newcommand{\secsum}{\vspace{0.25em}\textbf{\textit{Summary: }}} 

\secsum The overall model error observed among all models is in range of 0\%-0.089\% which shows that a learning model can be successfully generated using learned knowledge from similar applications. 
Moreover, the results confirm the applicability of knowledge transfer approaches in a cross application scenario. In addition, teh results also justify the feasibility of SVR, MPR and RF as base-learners to generate a learning model under \thename.

\begin{table}[!tb]
\footnotesize
\centering
\caption{Scenario-1: \%RRMSE for cross-applications on EC2. Transparent cell values represent the model error using \thename where the source applications serve as a source of knowledge for the target application. Highlighted cell values represent model error for three application-specific base models.}
\begin{tabular}{rr|ccc}
 & & \multicolumn{3}{c}{\textit{TARGET}} \\
 && \textbf{smallpt} & \textbf{Vard} & \textbf{Item Rec.} \\
 \midrule
 \multirow{9}{*}{\rotatebox{90}{\textit{SOURCE}}} && \cellcolor{CornflowerBlue!50} MPR-B 0.369186& MPR- 13.3162 & MPR- 0.7119354\\
 &\textbf{smallpt} &\cellcolor{CornflowerBlue!50} SVR-B 0.37245 & SVR- 14.48187 & SVR- 0.7197244 \\
 && \cellcolor{CornflowerBlue!50} RF-B 0.408157 & RF- 13.41974 & RF- 0.6699096   \\
 \cline{3-5}
 && MPR- 0.3611638 &\cellcolor{CornflowerBlue!50} MPR-B 13.30073& MPR- 0.684563 \\
 &\textbf{Vard} & SVR- 0.3658111 & \cellcolor{CornflowerBlue!50} SVR-B 14.45976 & SVR- 0.71643316 \\
 && RF- 0.4978552 & \cellcolor{CornflowerBlue!50} RF-B 13.47269& RF- 0.6699096 \\
 \cline{3-5}
 && MPR- 0.3720985 & MPR- 13.30905 & \cellcolor{CornflowerBlue!50} MPR-B 0.709919\\
 &\textbf{Item Rec.} & SVR- 0.3748514 & SVR- 14.48974 & \cellcolor{CornflowerBlue!50} SVR-B 0.717718 \\
 && RF- 0.4978552 & RF- 13.41974 & \cellcolor{CornflowerBlue!50} RF-B 0.654587 \\
\bottomrule
\end{tabular}
\label{tab:perc-relative-mse-ec2}
\end{table}

\subsubsection{Scenario 2: Cross-provider} \label{sec:Evaluation 2}
The second evaluation aims to ascertain whether \thename is able to satisfy the objective of enhancing the learning efficiency in a cross-provider scenario, \ie where the same application needs can be run on on different cloud providers. This is represented by the following example scenario:
\begin{mybox}
A model for application A can be generated to predict its performance on cloud Y using the learned knowledge of the same application having a model to predict its performance on cloud X.
\end{mybox}

\begin{figure*}[th]
\centering
\begin{tabularx}{1\textwidth}{@{}ccc@{}}

\includegraphics[width=0.32\textwidth, trim=0cm 0.2cm 0.2cm 0.25cm, clip]{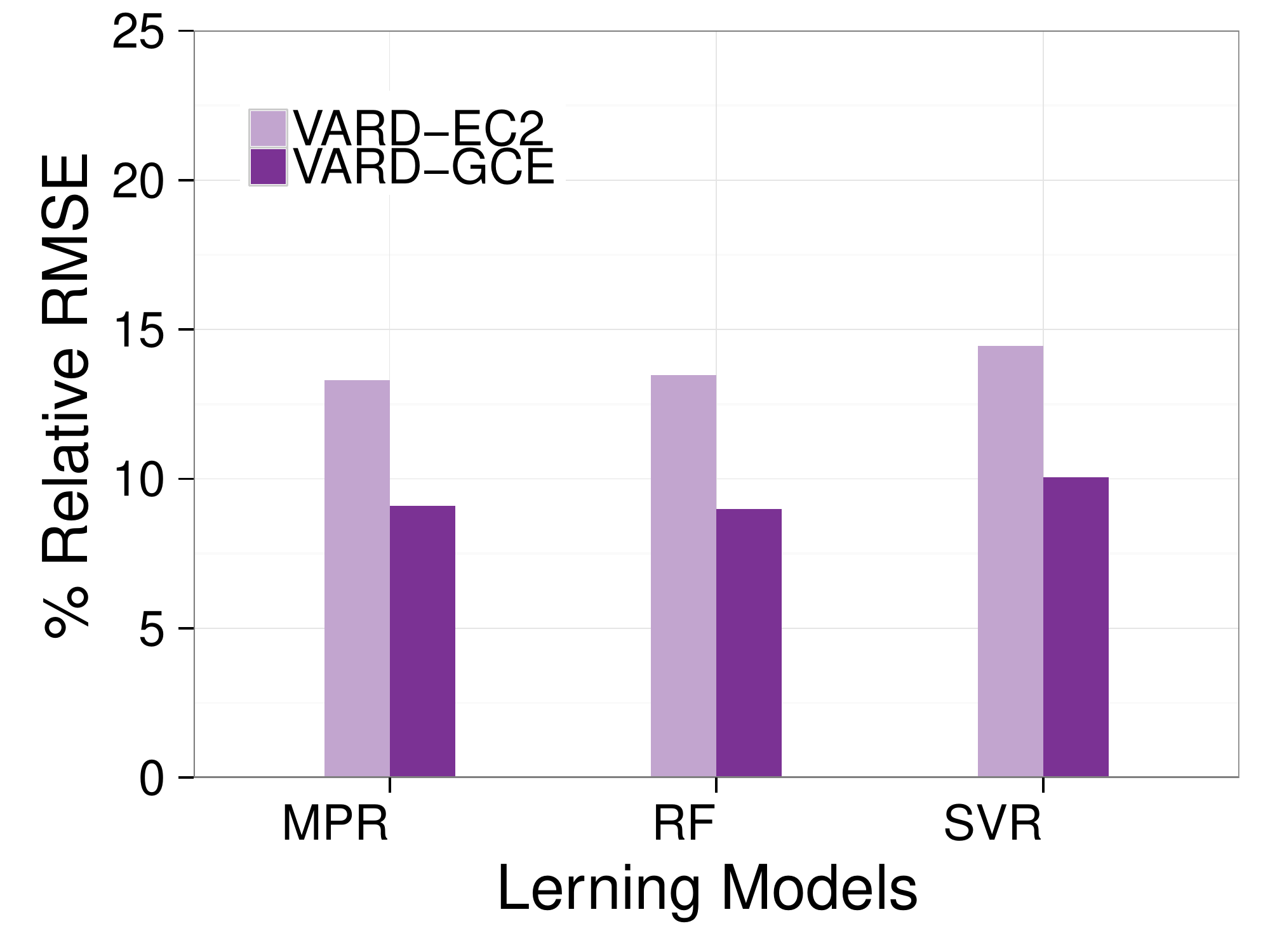}
&
\includegraphics[width=0.32\textwidth, trim=0cm 0.2cm 0.2cm 0.25cm, clip]{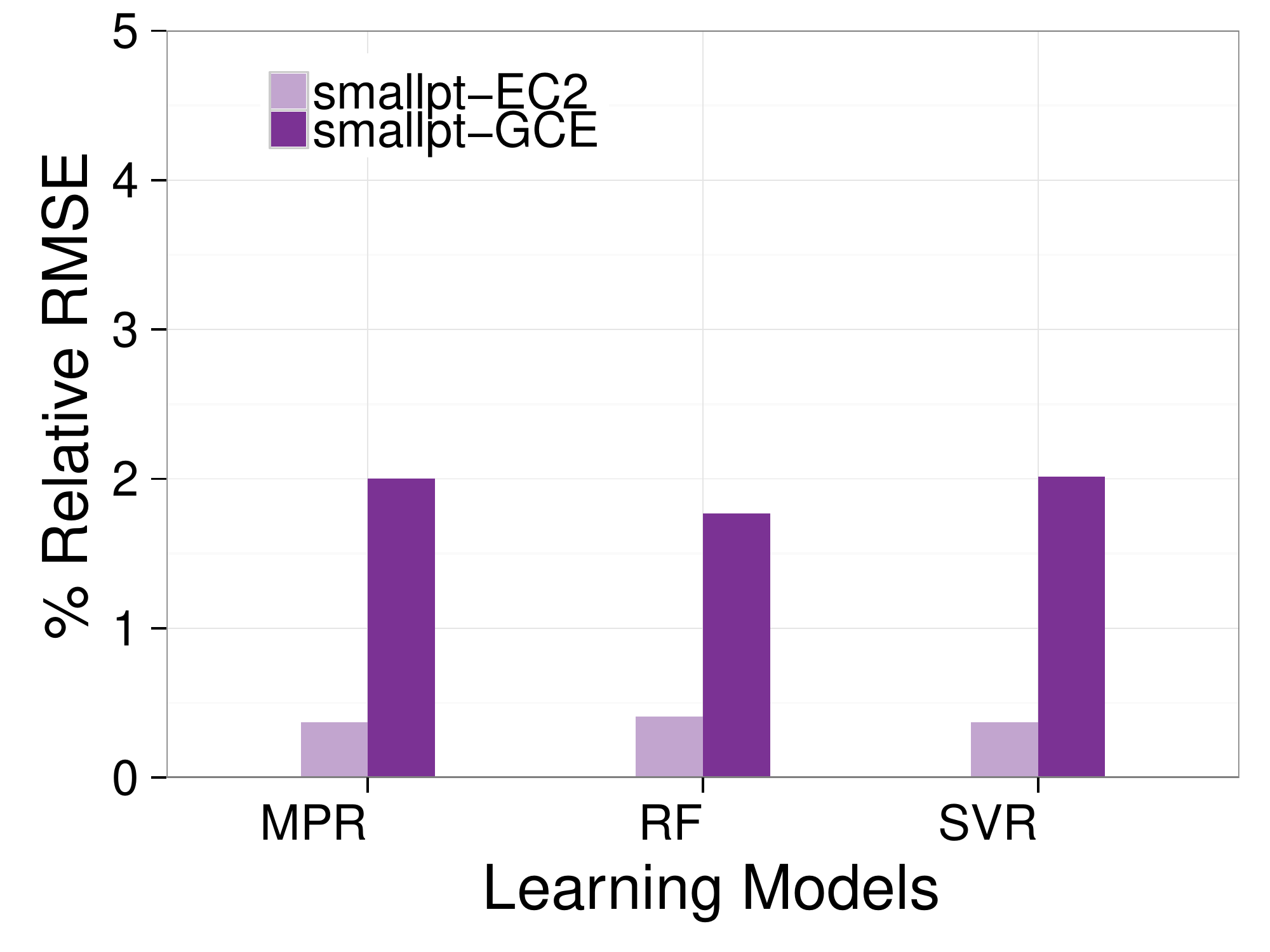}
&
\includegraphics[width=0.32\textwidth, trim=0cm 0.2cm 0.2cm 0.25cm, clip]{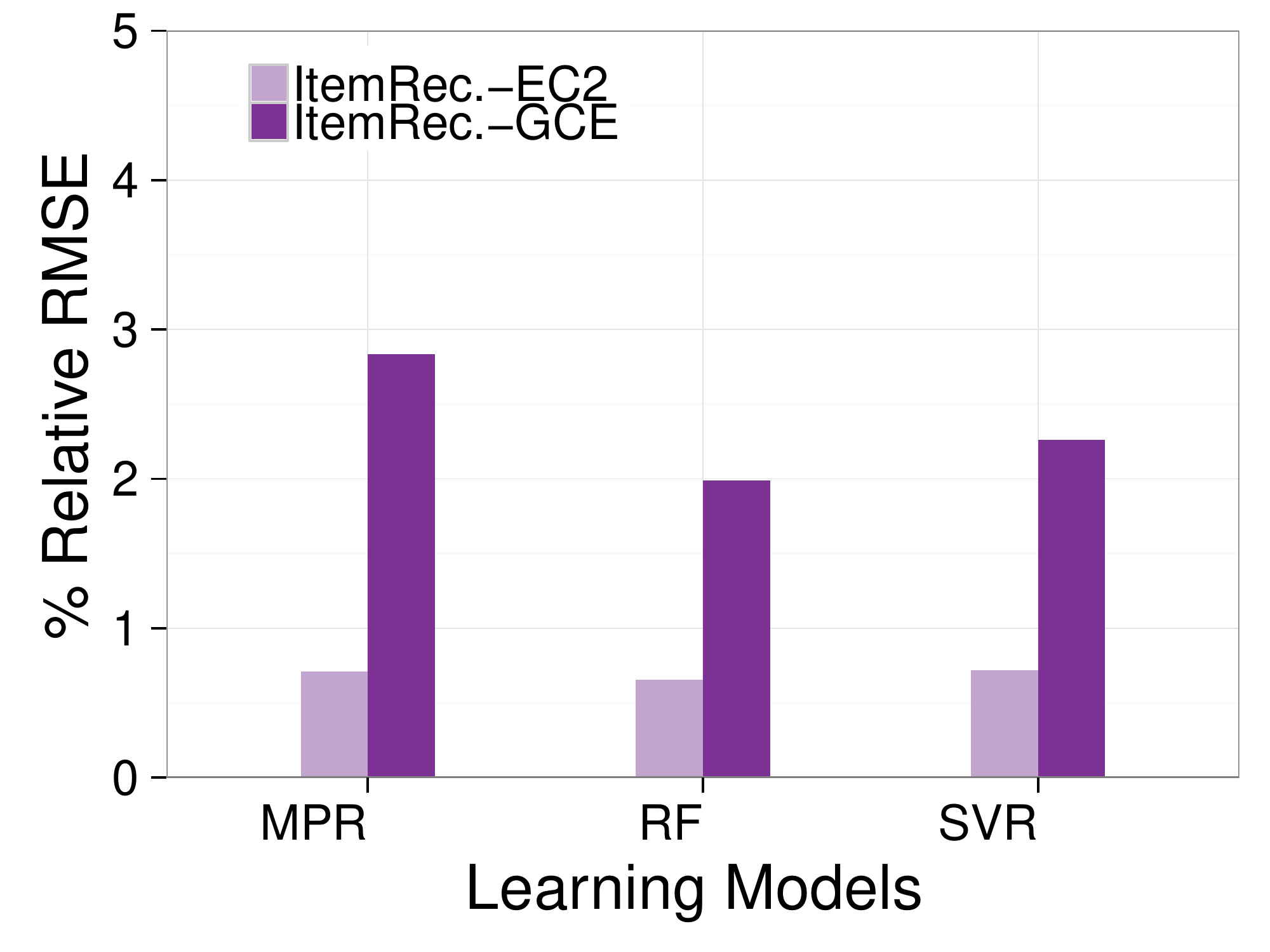}
\\
(a) & (b) & (c)
\end{tabularx}

\caption{Scenario-2: Model error for (a) Vard, (b) smallpt, and (c) ItemRecommender. Learning models are aligned on x-axis and each set of bars represents the error of applying the same model on different cloud providers.}
\label{fig:cross provider}
\end{figure*}

\begin{figure}[th]
\centering
\includegraphics[width=0.33\columnwidth, trim=0.2cm 0.2cm 0.2cm 0.2cm, clip]{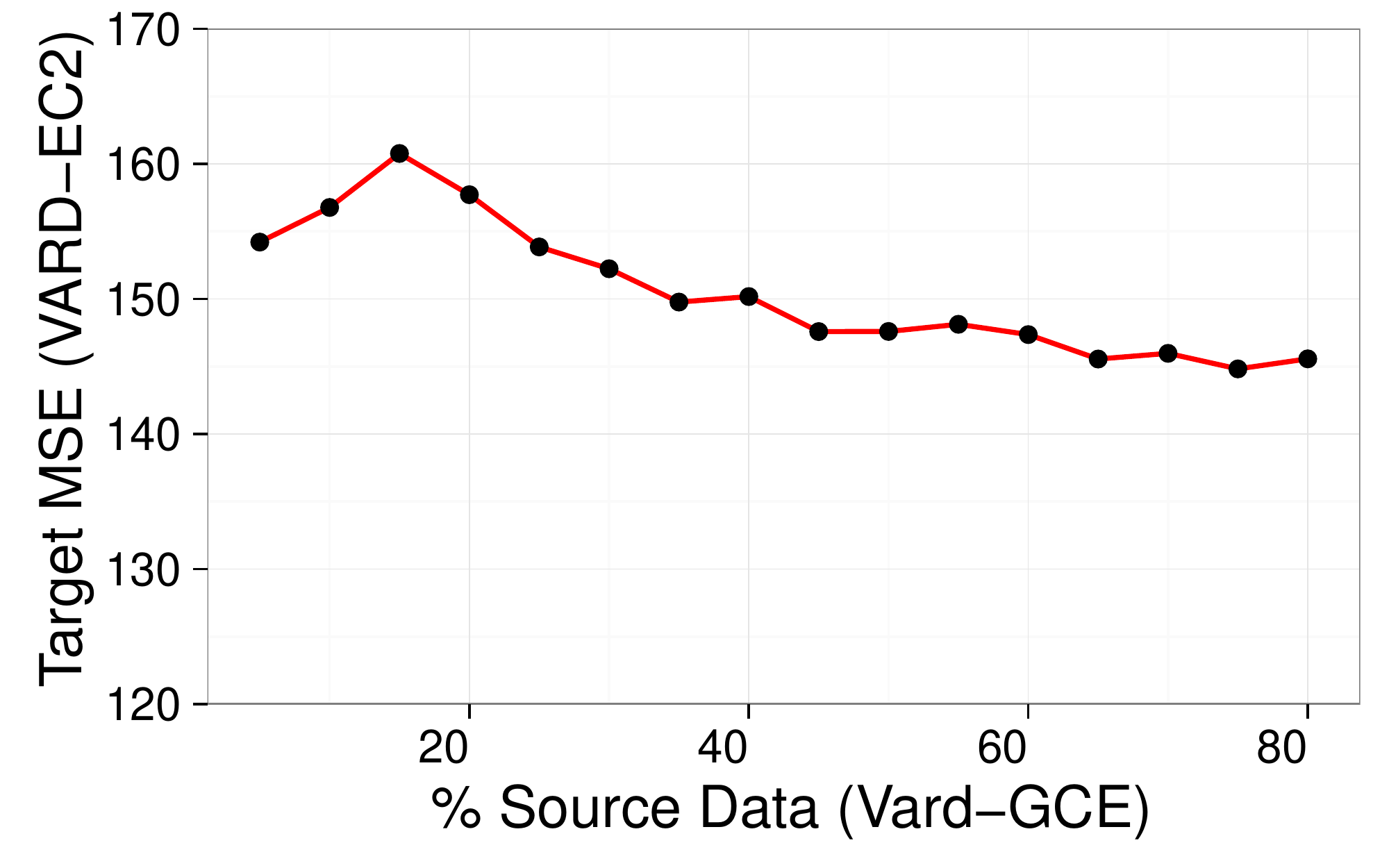}%
\includegraphics[width=0.33\columnwidth, trim=0.2cm 0.2cm 0.2cm 0.2cm, clip]{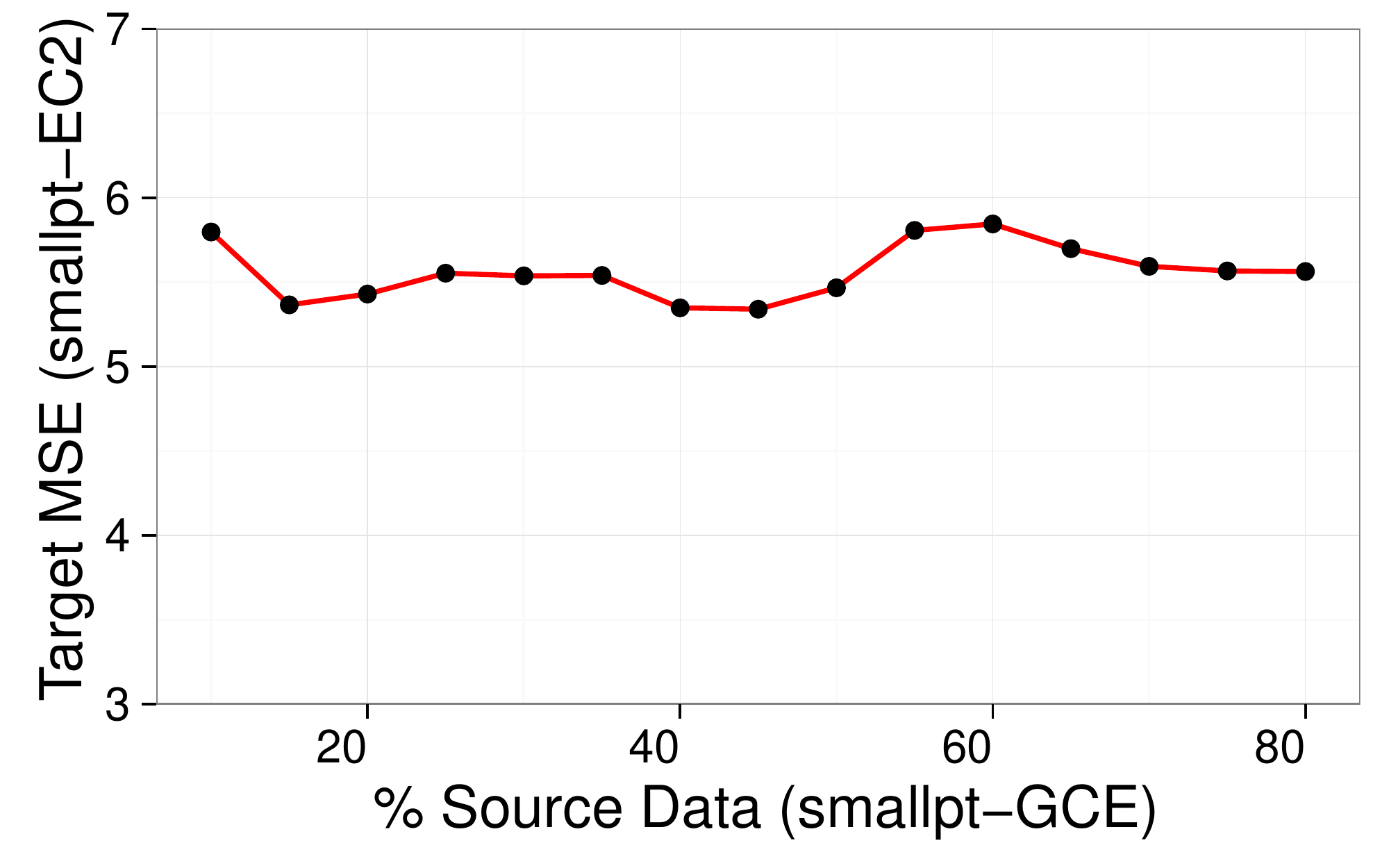}
\caption{Scenario-2: The effect of instance knowledge from a source in improving the model in a target domain (\ie reducing its MSE).}
\label{fig:reduced MSE}
\end{figure}

The data collected from the two cloud providers involved -- EC2 (source) and GCE (target) -- differ in feature space due to varying configuration settings at the VM level. 

Each plot in \fig{fig:cross provider} presents the model performance of each application on GCE where the same application's data on EC2 is used as the source of knowledge by \thename. Moreover, these plots show the associated error rate of the transferred model. According to the similarity outcome, each source application in each plot is tagged as closely similar and is applicable for the Transfer-All mode with SVR as base-learner. The left plot in \fig{fig:cross provider} clearly depicts a lower model error for Vard-GCE for SVR, MPR and RF where \%RRMSE for SVR model on EC2 is 14.459\% which is comparatively higher than 10.049\% on GCE. Similarly, the \%RRMSE for MPR and RF on EC2 is 13.300\%-13.472\%, a little higher than 9.094\%-8.995\% on GCE. 
In the middle plot, MPR, SVR and RF models are performing almost the same and interestingly with a model error of less than 2\% where model error for SVR, MPR and RF on EC2 is 0.3724\%, 0.3691\% and 0.4081\%, respectively. The models are evidently performing well on GCE where the \%RRMSE for SVR, MPR and RF is 2.0152\%, 2.001\% and 1.770\%, respectively.

The right plot shows similar behaviour where the model error for the ItemRecommender on GCE is a little higher for all base learners in \thename. The \%RRMSE for the SVR model on GCE is 2.262\% which is higher than 0.7177\% on EC2. A similar trend is seen with the MPR and RF models, where the model error on EC2 is 0.7099\%-0.654\% which is a little lower than 2.836\%-1.987\% on GCE.

Furthermore, there are some surprising findings: \thename results in significant reduction of model error compared to base models. For this result, the model accuracy metric is MSE as we are presenting diagnostics for one model with a particular data set. The reduced MSE is observed in two cases. First, this is observed when Vard-GCE instance knowledge is used as part of Transfer-All mode to generate a prediction model for Vard-EC2. SVR is used as the base-learner and Vard-GCE instance data is used as source knowledge. The top plot in \fig{fig:reduced MSE} describes the effect of instance data of source application (Vard-GCE) at the model training for predicting Vard-EC2 performance. The horizontal axis indicates the percentage use of source instance data mixed with the auxiliary target data, and the y-axis indicates the MSE value of test data. The MSE value is observed to consistently decrease on each percentage increase of source data. The MSE value of the base model is 177.960, which is far higher than the values seen in the graph. The lowest MSE by \thename is 144.822 for Vard-EC2. 

In the second case, a reduction in MSE is observed when smallpt-GCE instance knowledge is used as part of Transfer-All mode, with SVR as the base-learner, to generate a prediction model for smallpt-EC2. The results are presented in the bottom plot of \fig{fig:reduced MSE}. As opposed to the above trend, we do not observe a continuous reduction of MSE. Nonetheless, some lower bounds are observed. Most importantly, the use of source instance data is able to maintain a continuous reduced MSE compared to 9.746 as the base model MSE.

\secsum The results confirm the positive influence of transferred knowledge, validating the assumption of applying a TL approach. The experiment also shows that it is possible and feasible to make use of existing knowledge to increase model accuracy. The applicability of this approach is clearly evident in this cross-provider scenario. 
Furthermore, the reduced model error shows the sensitivity of prediction models to capture data variation on different distributional data.

\subsubsection{Scenario 3: Cross-application \& Cross-provider} \label{sec:Evaluation 3}
The final evaluation aims to identify whether using \thename helps in providing deployment decisions across different applications \emph{and} different cloud providers: in other words, is \thename able to enhance learning efficiency in a cross-application, cross-provider scenario where both (a) source and target applications are different, and (b) source and target providers are different. This is illustrated by the following scenario:
\begin{mybox}
A prediction model can be generated for the performance of application B on cloud Y using the learned knowledge of the performance of application A on cloud X.
\end{mybox}

Again, this evaluation involves EC2 and GCE. Source knowledge is transferred based on the fact that there is little or no similarity neither at application nor cloud provider level. 

Table \ref{tab:perc-relative-mse-GCE} summarizes the results when applications on GCE are using existing knowledge of different applications on EC2. The results for cross-application cases (highlighted in grey) are excluded from this table as already discussed in \ref{sec:Evaluation 2}. Each cell value in the table illustrates the \%RRMSE value for each model and is assessed on the test data from target application only. Under this scenario, the models are compared with each other to assess the model error for different base learners as well as evaluated for prediction accuracy on target application in comparison of the source application.

Considering smallpt-GCE as a target with Vard-EC2 (c1r2) and ItemRecommender-EC2 (c1r3) as source the percentage difference of model error for MPR-based models is 0.01\%, for SVR-based models the difference is 0.003\% and for RF-based models the error difference is 0.092\%. Similarly, Vard-GCE as the target application has the prediction models having a model error difference of 0.060\%, 0.028\% and 0.263\% for MPR, SVR, and RF, respectively. It can also be observed that the models generated for Vard-EC2 are presenting a better performance for Vard-GCE with reduced \%RRMSE.
The right most column displays values for ItemRecommender-GCE, where Vard-EC2 and smallpt-EC2 are considered as the source of knowledge. The \%RRMSE difference for MPR, SVR and RF-based models is 0.085\%, 0.055\%, and 0.158\%, respectively.
This shows that \thename is able to achieve reasonably accurate model generation using knowledge of a source domain and task that have no similarity at the cloud or application level. 
This also shows that a good feature representation can reduce domain differences even if the domains have some heterogeneity at the cloud provider level.

\secsum Perhaps not surprisingly, transferring knowledge across both application and cloud provider differences is not always successful however. Hence, this particular application of transfer knowledge is an area of future research.

\begin{table}[!tb]
\footnotesize
\centering
\caption{Scenario-3: \%RRMSE for cross-provider, here target applications are listed in columns and source in rows. Each cell value represents the model error for three models (MPR, SVR and RF) when the source domain is EC2 and served as a source of knowledge for model generation on GCE (target domain).}
\begin{tabular}{rr|ccc}
 & & \multicolumn{3}{c}{\textit{TARGET}} \\
 & & \textbf{smallpt-GCE} & \textbf{Vard-GCE} & \textbf{Item Rec.-GCE}  \\
 \midrule
 \multirow{9}{*}{\rotatebox{90}{\textit{SOURCE}}} & & \cellcolor{black!75}& MPR- 9.094947 & MPR- 2.836855 \\
 & \textbf{smallpt-EC2} &\cellcolor{black!75} & SVR- 10.04975 & SVR- 2.26189 \\
 & & \cellcolor{black!75}& RF- 8.995541 & RF- 1.9887718   \\
 \cline{3-5}
 & & MPR- 2.011354 &\cellcolor{black!75} & MPR- 2.75103 \\
 & \textbf{Vard-EC2} & SVR- 2.011354 & \cellcolor{black!75} & SVR- 2.31769 \\
 & & RF- 1.862442 & \cellcolor{black!75} & RF- 2.147283 \\
 \cline{3-5}
 & & MPR- 2.001243  & MPR- 9.161341 & \cellcolor{black!75}\\
 & \textbf{Item Rec.-EC2} & SVR- 2.015265  & SVR- 10.07811 & \cellcolor{black!75} \\
 & & RF- 1.770105 & RF- 9.258656 & \cellcolor{black!75} \\
\bottomrule
\end{tabular}
\label{tab:perc-relative-mse-GCE}
\end{table}

\subsection{Time and cost effectiveness}

In addition to the above validation of our applied TL scheme, the approach is been evaluated for time and cost-effectiveness.
A key feature of \thename is the utilization of existing knowledge in the form of the learned model (Transfer-Model) and data instances (Transfer-All). This results in a reduction of time to generate a model from scratch. Additionally, utilizing existing data instances reduces the cost required to collect a large amount of data for a new application.

\begin{figure}[th]
\centering
\includegraphics[width=0.5\columnwidth]{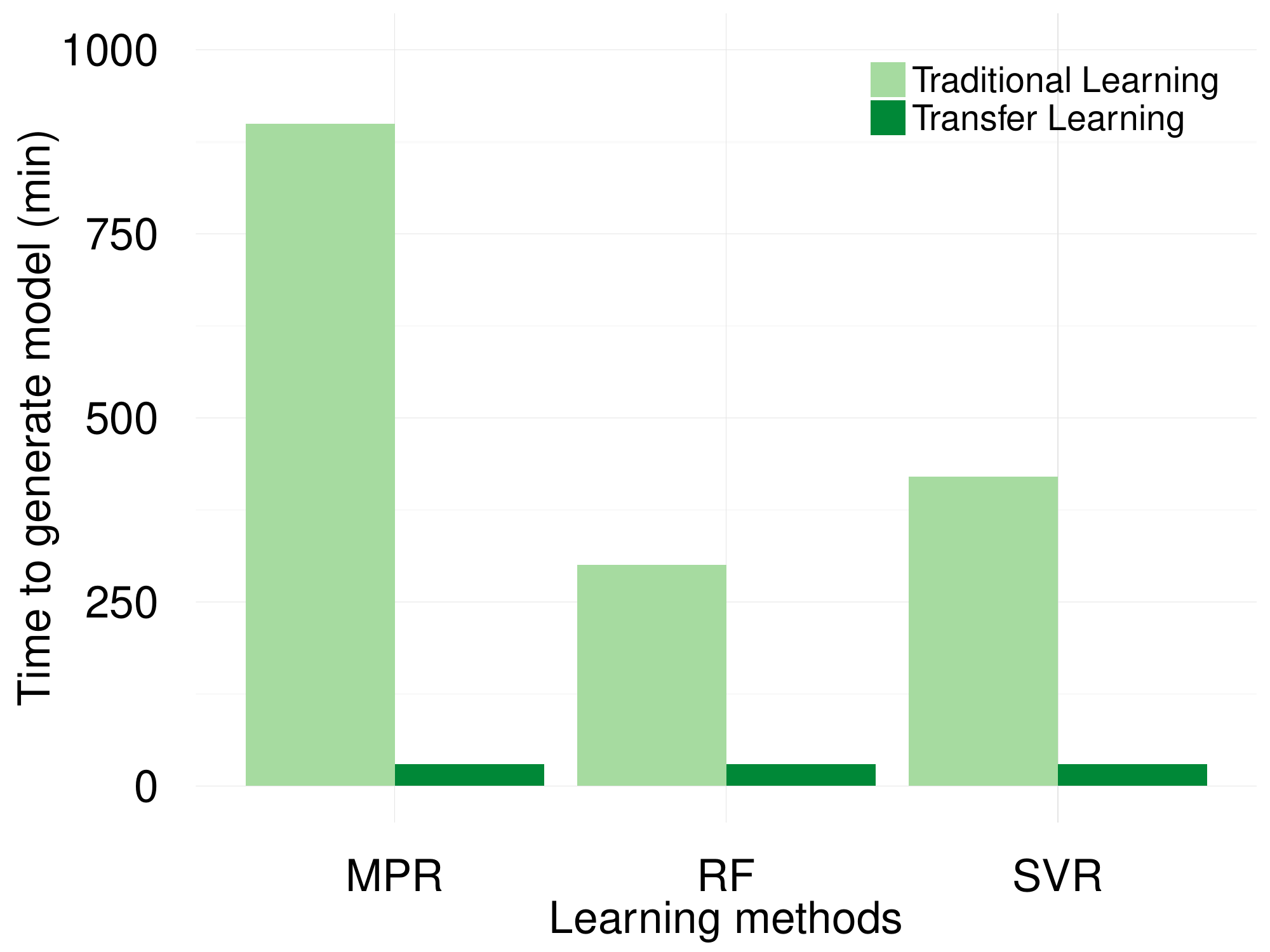}
\caption{The time taken to generate models using traditional ML and TL.}
\label{fig:cost and time}
\end{figure}

Model generation on its own incurs a variable amount of time at each stage: data pre-processing, feature selection, model calibration, parameter tuning, model training, and assessment~\cite{elkhatib2017diverseDS}. 
\fig{fig:cost and time} illustrates the approximate time needed to generate a learning model by using traditional model generation methods in comparison to transfer learning using our approach. 
MPR requires additional efforts to identify a best fit model and to examine non-linear curve fitting of multiple predictors with a response variable; \ie very \textit{transparent} in ML terms. Therefore, it incurs the highest time ($\approx$900 minutes) for model generation involving frequent human intervention to evaluate outcomes at each sub-stage.
SVR comes second ($\approx$ 420 minutes) where the majority of the time is used for parameter tuning. 
RF takes less time ($\approx$300 minutes), and is observed to offer better understanding of the relationship of predictors with the response variables as compared to SVR.

It is clearly seen from \fig{fig:cost and time} that TL can reduce the significant effort of model generation. The time for TL includes pre-processing of new data, similarity measurement, model transfer, training, assessment and any human intervention to check through the process.
Model accuracy is a proxy metric for model generation time and training cost: the lower the model error, the less time and cost required to train the new model as only a subset of the target application data is used to train the generated model. 
Such saving brought on by the use of existing knowledge is potentially very significant as collecting data to train a model is a lengthy and costly process in terms of costs needed for cloud resources to collect data samples. 
Overall, model training is made more efficient by saving 60.12\% of the required cost and time. 
Specifically, we were able to reduce VM usage for the purpose of data collection for a single application from 168 hours to 67 hours. 
This translates to saving \$92.33 out of \$153.88 on 8 EC2 VMs. Similarly, on GCE, a cost of \$53.323 from a total of \$88.872 was saved on one single experiment comprised of 7 nodes.

\section{Related Work}

A big challenge in cloud computing stems from the wide variety of technologies, APIs, and terminologies used \cite{elkhatib2016Crosscloud}. Furthermore, uncertainty associated with how these services are managed (\eg scheduling algorithms, load balancing policies, co-location strategies, 
\etc \cite{Maji2014interference,daleel,leitner2016patterns}) add a black-box effect to this complexity. 

Thus, providing customers with realistic and application-specific deployment decisions is an important and complex goal. In this decision process, application requirements are being matched with cloud resources. Broadly speaking, there are two approaches to such a decision-making process, which we review below.

\textbf{Metric-based solutions} rely on representing cloud resources and their capabilities in a certain way, such as using standardized KPIs (\eg~\cite{Garg@2013ranking,Gaurav@2016improvedrank,STRATOS,OPTIMIS}) 
or through benchmarking (\eg~\cite{Cloudcmp,Verghese2013Cloudbench}). 
The former method results, despite all efforts, in an outdated and reductive representation due to the sheer breadth and proliferation of the cloud computing market. 
The latter method avoids this through persistent benchmarking in an attempt to capture irregularities and attain a detailed and up-to-date performance profile for each different cloud resource type. This of course comes at a high operational cost. 
Moreover, a disadvantage of both methods is that they are based on application-agnostic ranking and not on knowing how the application will \textit{actually} perform on a given infrastructure.

\textbf{Application models} focus more on the other part of the matchmaking decision process: application requirements. Examples include vendor-independent ontologies (\eg~\cite{Petcu@2011mOSAIC,Kamateri2013cloud4soa}) and model-driven engineering (\eg~\cite{Balis2016PaaSage,menychtas2014software,jumagaliyev2019cadaml}). 
These solutions are heavily dependent on fine-grained information from domain experts, analysts and decision makers to get complete knowledge of business models and company strategies~\cite{jumagaliyev2017dsml}. 
As such, a designer must be aware of the impact of decisions, alternative decisions, actor interactions, dependencies, and processes while designing workflows and architectural models. Such processes require significant developer experience and time to follow domain-specific design principles. 

\textbf{ML-based methods} are also starting to emerge. Such methods aim to gain the best of both worlds by using experiential data (as benchmarks do) to model application behaviour on different deployment setups. 
A prominent trend in such studies is to focus on data analytics and MapReduce-style applications~\cite{Wieder2012Conductor, Gencer2015ResponseSurface,Huang2013ML,Chiang2014Matrix}, due to their operational footprint and having a recurrent workload pattern which is relatively easy to model. 
However, a common overhead here is training: significant data and time (which translates to cloud costs) are needed to train a model. Therefore, more recent efforts focus on reducing the learning cost. Ernest~\cite{Venkataram16Ernest} offers a performance prediction for analytic applications, and uses an optimal experimental design technique to select useful training point for model generation. This, however, requires extensive evaluation to optimize the cost-performance trade-off.  CherryPick~\cite{AlicherryPick2017} uses optimization and encoding methods to obtain a near optimal solution and to narrow down the search space. However, the solution is designed for recurrent analytic applications and is restricted to provide only one solution as an optimal choice. \thename, on the other hand, is not designed for any particular application domain and can optimize the cost-performance trade-off for different ranges of VMs. 
Paris~\cite{Yadwadkar2017paris} uses hybrid offline benchmarking to generate sufficient workload fingerprints to obtain a cost-performance trade-off. Though, combining synthetically generated data with real-world data will typically results in the reduction of prediction accuracy as is evident from the low prediction accuracy reported in the Paris paper as compared to that of \thename. 

\textbf{A final note:} Several recent works used ML to look into the variation of auctioned cloud resources (as opposed to on-demand ones), namely AWS Spot, \eg~\cite{kilcioglu2015revenue,Zheng2015bid,Wang2017spot}.
However, this has been proven to be a relatively trivial optimization problem \cite{Sharma2016} and is not of interest to our work.

\section{Conclusion}
Decision making in cloud environments is a challenging task due to the wide and ever expanding variety of IaaS service offerings. A customer entering such diverse market is likely to be confused with the range of choices on offer and without much knowledge about the selection criteria. A decision support system supplemented with traditional machine learning methods is therefore a very attractive service for cloud users. However, this inevitably comes with a significant learning time and monetary cost for making decisions specific to each application and cloud infrastructure. 

In this paper we present \thename, a novel solution to increase the efficiency of ML-assisted DSS to make application-specific decisions in a cross-cloud environment. The solution makes use of existing knowledge, transferring such knowledge to be used for other applications and/or cloud infrastructures.
More specifically, this work applies TL to identify the type of knowledge to be transferred and details a methodology to identify similarity between different sources of knowledge. 
\thename is not using source pre-trained models. Instead, it is training a function over the source and target application data. This approach is evaluated from three perspectives involving two public cloud providers and three applications of varying architectures. The evaluation results are very promising and identify a significant reduction in model generation overhead (of 60\%) in terms of time and cost. 
Thus, \thename is able to enhance the capability of intelligent decision support systems to make them more cost-effective. This important contribution is also applicable for multi-cloud brokers~\cite{Elhabbash2019brokersurvey} in considering a large amount of deployment options.
 
This work opens up a number of avenues for future research. In particular, it would be interesting to enrich the framework with supplementary learning methods including exploration of unsupervised transfer learning techniques~\cite{Usama2019unsup}. Moreover, it would be interesting to extend the study to deal with multi-criteria decision making and to better understand the generality of this approach by expanding to other categories of applications and cloud providers.

\bibliographystyle{abbrv}
\bibliography{cloudtl}

\end{document}